\documentclass[aps,twocolumn,showpacs,showkeys,preprintnumbers,amsmath]{revtex4}

\allowdisplaybreaks[1]
\usepackage{amsmath}
\usepackage{amssymb}
\usepackage{cases}
\usepackage{bm}
\usepackage{graphicx}
\usepackage{float}
\usepackage{isorot}

\newcommand{\nc}{\newcommand}
\nc{\rnc}{\renewcommand}
\nc{\dmo}{\DeclareMathOperator}

\nc{\nojour}[3]{\textbf{#1}, #2 (#3)}
\nc{\arXiv}[3]{arXiv:#1.#2v#3}
\nc{\ibid}[3]{{\em ibid.} \textbf{#1}, #2 (#3)}
\nc{\EPJB}[3]{Eur.\ Phys.\ J.\ B \textbf{#1}, #2 (#3)}
\nc{\EPL}[3]{Europhys.\ Lett.\ \textbf{#1}, #2 (#3)}
\nc{\JPA}[3]{J.\ Phys.\ A \textbf{#1}, #2 (#3)}
\nc{\JPAMG}[3]{J.\ Phys.\ A: Math.\ Gen.\ \textbf{#1}, #2 (#3)}
\nc{\JPAMT}[3]{J.\ Phys.\ A: Math.\ Theor.\ \textbf{#1}, #2 (#3)}
\nc{\JPC}[3]{J.\ Phys.\ C \textbf{#1}, #2 (#3)}
\nc{\JPCS}[3]{J.\ Phys.\ Conf.\ Ser.\ \textbf{#1}, #2 (#3)}
\nc{\JSP}[3]{J.\ Stat.\ Phys.\ \textbf{#1}, #2 (#3)}
\nc{\PA}[3]{Physica A \textbf{#1}, #2 (#3)}
\nc{\PR}[3]{Phys.\ Rev.\ \textbf{#1}, #2 (#3)}
\nc{\PRB}[3]{Phys.\ Rev.\ B \textbf{#1}, #2 (#3)}
\nc{\PRE}[3]{Phys.\ Rev.\ E \textbf{#1}, #2 (#3)}
\nc{\PRL}[3]{Phys.\ Rev.\ Lett.\ \textbf{#1}, #2 (#3)}
\nc{\RMP}[3]{Rev.\ Mod.\ Phys.\ \textbf{#1}, #2 (#3)}

\dmo{\sgn}{sgn}
\dmo{\diag}{diag}

\nc{\Ip}{I^{(+)}}
\rnc{\Im}{I^{(-)}}
\nc{\Ipm}{I^{(\pm)}}
\nc{\Ipsm}{I^{(+/-)}}
\nc{\If}{I^{(f)}}

\nc{\ba}{\begin{array}}
\nc{\ea}{\end{array}}
\nc{\bpi}{\begin{picture}}
\nc{\epi}{\end{picture}}
\nc{\bse}{\begin{subequations}}
\nc{\ese}{\end{subequations}}

\nc{\nn}{\nonumber}
\nc{\ds}{\displaystyle}
\nc{\ts}{\textstyle}
\nc{\scs}{\scriptstyle}
\nc{\unit}{\pmb{\openone}}
\nc{\p}{\partial}
\nc{\ua}{\uparrow}
\nc{\da}{\downarrow}
\nc{\uada}{{\uparrow\downarrow}}
\nc{\f}{\frac}
\nc{\fr}[2]{{\ts\f{#1}{#2}}}
\nc{\ph}{\phantom}

\nc{\al}{\alpha}
\nc{\be}{\beta}
\nc{\ga}{\gamma}
\nc{\de}{\delta}
\nc{\ep}{\epsilon}
\nc{\vp}{\varphi}
\nc{\la}{\lambda}
\nc{\om}{\omega}
\rnc{\th}{\theta}
\nc{\vt}{\vartheta}
\nc{\De}{\Delta}
\nc{\Th}{\Theta}

\nc{\Lbar}{\bar{L}}
\nc{\Ab}{\mathbf{A}}
\nc{\Abar}{\bar{\mathbf{A}}}
\nc{\Lb}{\mathbf{L}}
\nc{\Mb}{\mathbf{M}}
\nc{\Rb}{\mathbf{R}}
\nc{\Sb}{\mathbf{S}}
\nc{\Ub}{\mathbf{U}}
\nc{\xb}{\mathbf{x}}
\nc{\Xib}{\mathbf{\Xi}}
\nc{\nhb}{\mathbf{\hat{n}}}
\nc{\lab}{\bm{\la}}

\nc{\Acal}{{\cal A}}
\nc{\Acald}{{\cal A}_d}
\nc{\Fcal}{{\cal F}}
\nc{\FcalCasiso}{{\cal F}_\text{Cas,iso}}
\nc{\Gcal}{{\cal G}}
\nc{\Gcaliso}{{\cal G}_\text{iso}}
\nc{\Acaliso}{{\cal A}_\text{iso}}
\nc{\Gcald}{{\cal G}_d}
\nc{\Xcal}{{\cal X}}
\nc{\Ocal}{{\cal O}}
\nc{\Xiso}{X_\text{iso}}

\nc{\fed}[1]{f_\text{#1}}
\nc{\fedb}{f_\text{b}}
\nc{\fedh}[1]{\hat{f}_\text{#1}}
\nc{\fedt}[1]{\tilde{f}_\text{#1}}

\nc{\xt}{\tilde{x}}
\nc{\xbar}{\bar{x}}
\nc{\N}{N}
\nc{\Lp}{L_\parallel}
\nc{\Np}{N_\parallel}
\nc{\Tc}{T_\text{c}}

\nc{\xig}{\xi_>}
\nc{\xik}{\xi_<}
\nc{\xipz}{\xi_{+,0}}
\nc{\ximz}{\xi_{-,0}}
\nc{\xipmz}{\xi_{\pm,0}}
\nc{\ar}{r}
\nc{\Vco}{V_\text{co}}
\nc{\Asite}{A_\text{site}}

\nc{\bc}{{bc}}
\nc{\pbc}{\text{(p)}}
\nc{\abc}{\text{(a)}}
\nc{\psabc}{\text{(p/a)}}
\nc{\zzbc}{\text{(00)}}
\nc{\ppbc}{\text{(++)}}
\nc{\zpbc}{\text{(0+)}}

\begin{document}

\bibliographystyle{apsrev}

\title{Universal anisotropic finite-size critical behavior of the
two-dimensional Ising model on a strip and of $d$-dimensional models on
films}
\author{Boris Kastening}
\email[Email address: ]{bkastening@matgeo.tu-darmstadt.de}
\affiliation{Institute for Materials Science,
Technische Universit\"at Darmstadt, 64287 Darmstadt, Germany}
\date{\today}

\begin{abstract}
Anisotropy effects on the finite-size critical behavior of a
two-dimensional Ising model on a general triangular lattice in an
infinite-strip geometry with periodic, antiperiodic, and free boundary
conditions (\bc) in the finite direction are investigated.
Exact results are obtained for the scaling functions of the finite-size
contributions to the free energy density.
With $\xig$ the largest and $\xik$ the smallest bulk correlation length
at a given temperature near criticality, we find that the dependence of
these functions on the ratio $\xik/\xig$ and on the angle parameterizing
the orientation of the correlation volume is of geometric nature.
Since the scaling functions are independent of the particular microscopic
realization of the anisotropy within the two-dimensional Ising model,
our results provide a limited verification of universality.
We explain our observations by considering finite-size scaling of free
energy densities of general weakly anisotropic models on a $d$-dimensional
film, i.e., in an $L\times\infty^{d-1}$ geometry, with {\bc} in the finite
direction that are invariant under a shear transformation relating the
anisotropic and isotropic cases.
This allows us to relate free energy scaling functions in the presence of
an anisotropy to those of the corresponding isotropic system.
We interpret our results as a simple and transparent case of anisotropic
universality, where, compared to the isotropic case, scaling functions
depend additionally on the shape and orientation of the correlation
volume.
We conjecture that this universality extends to cases where the geometry
and/or the {\bc} are not invariant under the shear transformation and
argue in favor of validity of two-scale factor universality for
weakly anisotropic systems. 
\end{abstract}



\maketitle

\section{Introduction}
\label{intro}
Bulk critical phenomena can be divided into distinct universality classes
\cite{Fi74}.
Critical exponents, certain critical amplitude ratios, and the
near-critical behavior of thermodynamic functions are identical for the
members of such a universality class and are called universal, since they
depend only on macroscopic properties such as near-critical correlation
lengths and not on the microscopic details of the system under
consideration.
For instance, the correlation length in the asymptotic critical domain,
i.e., for asymptotically small positive or negative $t\equiv(T-\Tc)/\Tc$,
is, for isotropic systems,
described by
\begin{align}
\label{xi.crit}
\xi&=\xipmz|t|^{-\nu}, & T&\gtrless\Tc,
\end{align}
where $\Tc$ is the bulk critical temperature, $\nu$ is a universal
critical exponent, $\xipmz$ are nonuniversal critical amplitudes, and
$R_\xi\equiv\xipz/\ximz$ is a universal critical amplitude ratio.

An extension to the concept of bulk universality concerns systems that are
geometrically confined on a length scale $L$ in one or more directions,
where $L$ is large compared to all microscopic length scales of the
system, such as lattice spacings (see \cite{Fi70,PrFi84,Pr88}; for
reviews see, e.g., \cite{Pr90,PrAhHo91}).
Consider the free energy density $f$ in units of $k_\text{B}T$ (this
normalization is used without further mentioning for all free energy
densities throughout this work) of a $d$-dimensional system that is
isotropic at such large distances.
Assume that $f$ may be split uniquely into a nonsingular and a singular
contribution according to
\begin{align}
\label{fns.fs}
f(T,L)=\fed{ns}(T,L)+\fed{s}(t,L),
\end{align}
where the singular contribution $\fed{s}$ is defined as that part of $f$
that becomes singular in $t$ at $t=0$ in the bulk limit $L\to\infty$.
If $f_s$ exhibits scaling, its behavior in the asymptotic critical domain
of large $L$ and small $|t|$, where Wegner corrections to scaling
\cite{We72} are negligible, may be described by a scaling function $\Fcal$
according to
\begin{align}
\label{Fcal}
L^df_s(t,L)=\Fcal(\xt),
\end{align}
with the scaling variable $\xt\equiv (L/\xipz)^{1/\nu}t$.
If $\Fcal$ exists, it is expected to be universal.
It describes the scaling behavior of the asymptotic singular part of the
free energy density for given system geometry and boundary conditions
(\bc) for the bulk universality class under consideration for the
isotropic case.

With the bulk free energy density
\begin{align}
\label{fb}
\fedb(T)\equiv\lim_{L\to\infty}f(T,L),
\end{align}
we may split the free energy density according to
\bse
\label{fex.fsf.ffs.def}
\begin{align}
f(T,L)&=\fedb(T)+\fed{ex}(T,L),
\\
\label{fsf.ffs.def}
\fed{ex}(T,L)&=\fed{sf}(T,L)+\fed{fs}(T,L),
\end{align}
\ese
where $\fed{ex}$ is called the excess free energy density,
$\fed{sf}(T,L)=L^{-1}\bar{f}_\text{sf}(T)$ represents any surface or
interface contributions, and $\fed{fs}$ is called the finite-size
contribution to the free energy density.
Denote the singular parts of  $\fed{ex}$, $\fed{sf}$, and $\fed{fs}$
by $\fed{ex,s}$, $\fed{sf,s}$, and $\fed{fs,s}$, respectively.
If the leading contributions to these singular parts exhibit
scaling, we may write in the asymptotic critical domain
\bse
\label{Acal.Gcal}
\begin{align}
\label{Acal}
L^d\fed{ex,s}(t,L)&=\Acal(\xt),
\\
\label{Gcal}
L^d\fed{fs,s}(t,L)&=\Gcal(\xt),
\end{align}
\ese
with scaling functions $\Acal$ and $\Gcal$.
These, again, are expected to be universal.
The Casimir amplitude is defined as the critical value of $\Gcal$, i.e.,
by
\begin{align}
\label{De}
\De&\equiv\Gcal(0).
\end{align}
The physical motivation for investigating the scaling functions $\Acal$
and $\Gcal$ and the Casimir amplitude lies in their close connection to
the critical Casimir force and its scaling function, see, e.g., the
monographs \cite{Kr94,BrDaTo00}, and the recent overview article
\cite{Ga09} (see also Appendix \ref{iso.scal.func}).
Ref.~\cite{Ga09} also provides a collection of recent theoretical and
experimental results.

Now consider weakly anisotropic systems, which are characterized by a
single bulk correlation-length exponent $\nu$.
We assume here that their bulk correlation lengths are related by a
shear transformation to the isotropic case.
Then $\nu$ assumes in any direction the same value as for the isotropic
case, which is supported by other investigations
\cite{Gr70,Br74,WuMCTrBa76,CaPeRoVi98,BlCoHo87}.
On the other hand, the $\xipmz$ depend on the direction and are, for a
$d$-dimensional system, described by the surface of a $d$-dimensional
ellipsoid.
However, in any given direction, Eq.\ (\ref{xi.crit}) holds and
$R_\xi$ assumes its isotropic value.

The situation is less clear for the scaling functions on the right hand
sides of Eqs.\ (\ref{Fcal}) and (\ref{Acal.Gcal}), since they may
additionally depend on the parameters describing the shape and the
orientation of the correlation lengths ellipsoid, i.e., on
$d{-}1$ correlation length ratios and $d(d{-}1)/2$ angles.
These additional $d(d{+}1)/2{-}1$ parameters may be organized into a
$d\times d$ symmetric matrix $\Abar$ with $\det\Abar=1$ \cite{ChDo04}.
The question arises whether such a function can be considered universal
\cite{Do06,Do08,DiCh09}.
Dohm \cite{Do08} interprets any dependence on $\Abar$ as nonuniversal.
Diehl and Chamati \cite{DiCh09} suggest to define universality only after
transforming to an isotropic system by means of a shear transformation.
Here we advocate the interpretation that scaling functions even for
weakly anisotropic systems are universal, if they depend on the anisotropy
only through its long-distance properties parameterized by $\Abar$ and not
on any microscopic details of how it is realized.
We will return to this issue in Sec.~\ref{summary}.

For simplicity, we consider only systems, where the transformation to an
isotropic system leaves the geometry and the {\bc} invariant.
In particular, we compute free energy scaling functions for the specific
case of an anisotropic two-dimensional Ising model on an infinite strip of
width $L$ and explain our explicit exact results by investigating systems
confined to $d$-dimensional films of width $L$.
We consider periodic, antiperiodic, fixed, and free {\bc} in the direction
of the width $L$ of the film.
For all cases we define a scaling variable by
\begin{align}
\label{xt}
\xt&\equiv(L/\xipz^{(L)})^{1/\nu}t,
\end{align}
where $\xipz^{(L)}$ is the $T>\Tc$ amplitude of the bulk correlation
length $\xi_L$ perpendicular to the film boundaries with a corresponding
asymptotic critical behavior
\begin{align}
\label{xi.crit.L}
\xi_L&=\xipmz^{(L)}|t|^{-\nu}, & T&\gtrless\Tc,
\end{align}
where $\xipz^{(L)}/\ximz^{(L)}=R_\xi$.

We do not consider complications arising from strong anisotropies
\cite{To07}, from subleading long-range interactions \cite{ChDo02,DiCh09},
or from scaling violations for large $\xt$ arising in a region of large
$L$ for fixed $\xi_L$ that manifest themselves in a nonuniform convergence
of the leading singular part of free energy densities towards the
respective scaling function in the asymptotic critical domain
\cite{ChDo99}.

This work is structured as follows.
In Sec.~\ref{I2.fe}, we discuss the two-dimensional Ising model on a
triangular lattice in infinite-strip geometry.
In Sec.~\ref{model}, the model is set up and basic quantities are defined.
In Sec.~\ref{corr.lengths.3c}, we recall explicit results for the bulk
correlation lengths and discuss their behavior near $\Tc$.
In Sec.~\ref{free.energy}, we derive the scaling behavior of the singular
contributions to $\fed{ex}$, $\fed{sf}$, and $\fed{fs}$ for periodic,
antiperiodic (Sec.\ \ref{free.energy.pa}), and free
(Sec.\ \ref{free.energy.00}) {\bc} to the extent they are defined.
We derive explicit expressions for the corresponding scaling functions and
critical Casimir amplitudes.
In Sec.~\ref{d-dim.film}, we discuss the finite-size scaling behavior for
free energy densities of $d$-dimensional anisotropic films with periodic
and antiperiodic {\bc} (Sec.\ \ref{d-dim.film.pa.bc}) and fixed and free
{\bc} (Sec.\ \ref{d-dim.film.fixed.bc}) and compare our results to those
found in Sec.\ \ref{free.energy}.
The discussion of the results of Secs.~\ref{I2.fe} and \ref{d-dim.film} is
delegated to Sec.~\ref{summary}.

\section{Two-dimensional Ising model}
\label{I2.fe}

\subsection{Basic definitions}
\label{model}
\begin{figure}
\begin{center}
\includegraphics[width=6.1cm,angle=0]{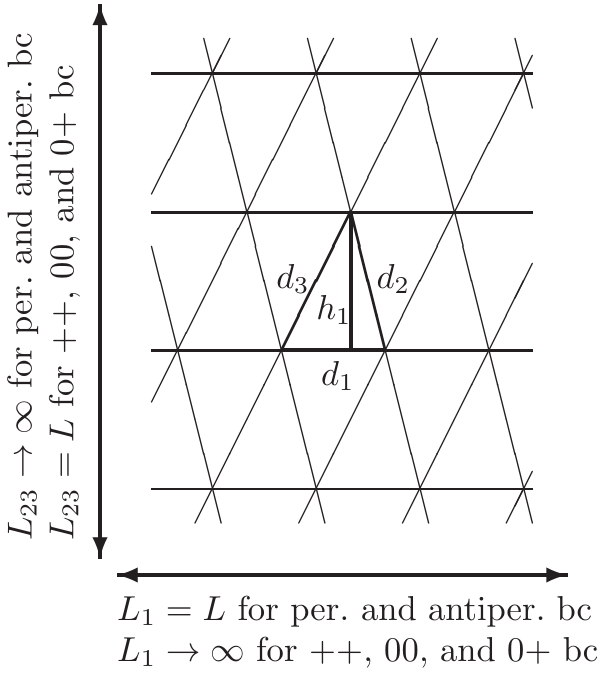}
\end{center}
\caption{\label{triang.lattice}
Triangular lattice with lattice constants $d_i$ on an infinitely long
strip of width $L$.}
\end{figure}

Consider a two-dimensional Ising model on a general triangular lattice
with lattice constants $d_i$ along the three lattice directions, see
Fig.~\ref{triang.lattice}.
For simplicity, we assume a ferromagnetic model with only nearest-neighbor
couplings $J_i$ along the sides of length $d_i$.
Consider this model on an $L_1\times L_{23}$ rectangular geometry, with
the $L_1$ direction parallel to the ``1'' lattice direction.
Let there be $\N_1$ layers in the ``1'' direction and $\N_{23}$ layers in
the ``23'' direction perpendicular to the ``1'' direction, so that
\bse
\begin{align}
\label{L1}
L_1&=\N_1d_1,\\
\label{L23}
L_{23}&=\N_{23}h_1,
\end{align}
\ese
where $h_1$ is the height of the elementary triangle with respect to the
side of length $d_1$, see Fig.~\ref{triang.lattice}.

Periodic or antiperiodic {\bc} are easiest implemented along one of the
three lattice directions.
We want this direction to be parallel to one of the rectangle edges and
therefore choose the ``1'' direction.
Periodic {\bc} are imposed by identifying a line of spins along the ``2''
direction (or, equivalently, the ``3'' direction) with spins along such a
line at a distance $L_1$ in the ``1'' direction (one may think of the
lattice as wrapped around a cylinder of circumference $L_1$, whose axis
points along the ``23'' direction).
For antiperiodic {\bc}, in addition the signs of the couplings $J_1$ along
one side of one such line are reversed.

Both free (``0'') {\bc}, i.e., no further neighboring spin for the last
spins on one edge of the rectangle, and fixed (``+'') {\bc}, i.e., a
fictitious neighboring spin with fixed value $+1$ for each last spin on
one edge of the rectangle, are easiest implemented if such an edge is
along one of the lattice directions.
For the rectangle defined above, this leads to the unique choice of 
imposing ``00,'' ``++,'' or ``0+'' {\bc} in the ``23'' direction, where
the two entries refer to the two opposite sides of the rectangle. 

With $\be\equiv1/(k_\text{B}T)$ and $K_i\equiv\be J_i$, the Hamiltonian of
this model reads
\begin{align}
\label{H}
-\be H
=
\sum_{m,n}\big(&K_1s_{m,n}s_{m+1,n}+K_2s_{m,n}s_{m,n+1}
\nn\\
&+K_3s_{m,n}s_{m+1,n+1}\big).
\end{align}
With the partition function
\begin{align}
\label{Z}
Z(T,L_1,L_{23})
&=
\sum_{\{s_{i,j}=\pm1\}}e^{-\be H},
\end{align}
the rectangle free energy density is given by
\begin{align}
f_\text{rect}(T,L_1,L_{23})
&=
-(L_1L_{23})^{-1}\ln Z(T,L_1,L_{23}).
\end{align}
We are interested in the free energy density
\begin{align}
\label{f}
f(T,L)\equiv\lim_{\Lp\to\infty}
\begin{cases}
f_\text{rect}(T,L,\Lp)
& \text{for per., antiper.~bc,}
\\[1ex]
f_\text{rect}(T,\Lp,L)
& \text{for ++, 00, 0+ \bc,}
\end{cases}
\end{align}
of an infinitely long strip, i.e., for an $L\times\infty$ geometry, with
periodic, antiperiodic, 00, ++, or 0+ {\bc} in the $L$ direction.
For the width $L$ of the strip holds
\bse
\begin{numcases}{L=}
\label{L.L1}
L_1 & \text{for periodic and antiperiodic \bc,}\qquad\qquad
\\[1ex]
\label{L.L23}
L_{23} & \text{for ++, 00, and 0+  \bc,}
\end{numcases}
\ese
and the same relation between $\N$, $\N_1$, and $\N_{23}$.

\subsection{Bulk correlation lengths}
\label{corr.lengths.3c}
\begin{figure}
\begin{center}
\includegraphics[width=6.2cm,angle=0]{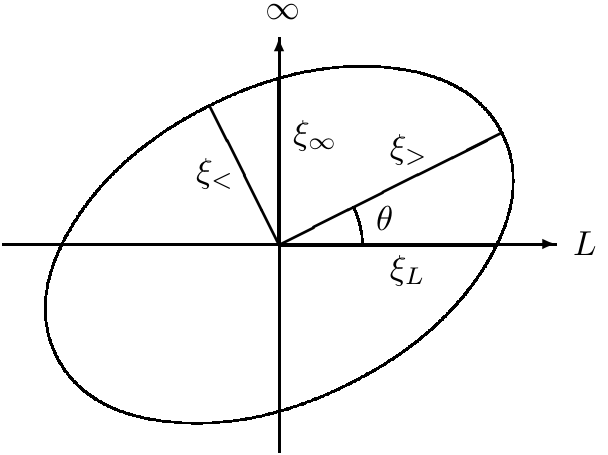}
\end{center}
\caption{\label{correlation.tilt}
Bulk correlation lengths ellipse with largest and smallest correlation
lengths $\xig$ and $\xik$, respectively. 
The $L$ direction is the direction perpendicular to the infinite direction
of the strip and $\xi_L$ and $\xi_\infty$ are the bulk correlation lengths
in these directions.}
\end{figure}

For two-dimensional systems, the ellipsoid of Sec.\ \ref{intro} describing
the bulk correlation lengths in the asymptotic critical domain reduces to
an ellipse, having a major radius $\xig$ and a minor radius $\xik$.
Define a bulk correlation ``volume'' by
\begin{align}
\label{A.co.def}
\Vco\equiv\xig\xik
\end{align}
and an aspect ratio by
\begin{align}
\label{r}
\ar&\equiv\xik/\xig, & 0&<r\leq1.
\end{align}
Let the direction of the major radius be rotated by an angle $\th$ with
respect to the $L$ direction of the infinitely long strip, see
Fig.~\ref{correlation.tilt}.

Call $\xi_\infty$ the bulk correlation length in the infinite-length
direction of the strip and $\xi_L$ the bulk correlation length in the
perpendicular direction.
Basic geometric considerations provide the relations
\bse
\label{xi.infty.L.2}
\begin{align}
\label{xi.infty.2}
\xi_\infty^{-2}&=\xig^{-2}\sin^2\th+\xik^{-2}\cos^2\th,
\\
\label{xi.L.2}
\xi_L^{-2}&=\xig^{-2}\cos^2\th+\xik^{-2}\sin^2\th,
\end{align}
\ese
between the various correlation lengths in the asymptotic critical domain.

Let $\xi_i$, $i=1,2,3$ be the bulk correlation lengths in the
$i$th lattice direction.
It follows that
\bse
\begin{numcases}{\xi_1=}
\label{xi1.xiL}
\xi_L & \text{for periodic and antiperiodic \bc,}\qquad\qquad
\\[1ex]
\label{xi1.xiinfty}
\xi_\infty & \text{for ++, 00, and 0+  \bc.}
\end{numcases}
\ese
Generalizing (\ref{xi.crit}), the bulk correlation lengths behave, in the
asymptotic critical domain, according to
\begin{align}
\label{xipm.i.t}
\xi_i&=\xipmz^{(i)}|t|^{-\nu} & T&\gtrless\Tc,
\end{align}
with $\xipz^{(i)}/\ximz^{(i)}=R_\xi$.
For the two-dimensional Ising model, well-known exact results are
\begin{align}
\label{nu.xi0p.xi0m}
\nu&=1, & R_\xi&=2.
\end{align}

According to Eq.~(A22) of Ref.~\cite{InNiWa86}, the $T>\Tc$ asymptotic
bulk correlation lengths are given by \cite{comment_InNiWa86_a}
\begin{align}
\label{xii}
d_i/\xi_i=-\ln\ga^{(i)},
\end{align}
where
\begin{align}
\label{ga.i}
\ga^{(i)}\equiv\f{a^{(i)}+\sqrt{{a^{(i)}}^2-4b^{(i)}c^{(i)}}}{2c^{(i)}},
\end{align}
with
\bse
\begin{align}
a^{(i)}&\equiv2z_i(1+z_j^2)(1+z_k^2)+4z_jz_k(1+z_i^2),
\\
b^{(i)}&\equiv z_i^2(1-z_j^2)(1-z_k^2),
\\
c^{(i)}&\equiv(1-z_j^2)(1-z_k^2),
\end{align}
\ese
with $i,j,k=1,2,3$ or cyclic permutations, and with
\begin{align}
\label{zi}
z_i\equiv\tanh K_i.
\end{align}
At the critical point the bulk correlation lengths diverge and therefore
$\ga^{(i)}=1$.
This condition may be written as
\begin{align}
\label{crit.point.2}
z_1+z_2+z_3+z_2z_3+z_3z_1+z_1z_2-z_1z_2z_3=1,
\end{align}
so that
\begin{align}
\label{crit.point.3}
z_{1,\text{c}}(z_2,z_3)=\f{1-z_2z_3-z_2-z_3}{1-z_2z_3+z_2+z_3}
\end{align}
is the critical value of $z_1$ for given $z_2$ and $z_3$.
Combining (\ref{xipm.i.t})--(\ref{ga.i}), we may expand around the
critical point through linear order in $t$ resulting in
\begin{align}
\label{z1.z1c.exp}
z_1=
z_{1,\text{c}}(z_2,z_3)
-\f{(1-z_2^2)(1-z_3^2)}{(1-z_2z_3+z_2+z_3)^2}
\f{d_1t}{\xipz^{(1)}}
\end{align}
and relate the $\xi_i$ according to 
\begin{align}
\label{xi.ratios}
\f{d_1/\xi_1}{2(1{-}z_2z_3)(z_2{+}z_3)}
&=
\f{d_2/\xi_2}{(1{-}z_2^2)(1{+}z_3^2)}
=
\f{d_3/\xi_3}{(1{-}z_3^2)(1{+}z_2^2)},
\end{align}
where terms of higher than linear order in $t$ have been omitted.

Geometric considerations, which we do not reproduce here, yield
\begin{align}
\label{A.co.res}
\Vco
&=
\f{A_\text{tr}(d_1,d_2,d_3)}{A_\text{tr}(d_1/\xi_1,d_2/\xi_2,d_3/\xi_3)},
\end{align}
where the function
\begin{align}
\label{A.triangle}
A_\text{tr}(\ell_1,\ell_2,\ell_3)
=\f{1}{4}\sqrt{2(\ell_1^2\ell_2^2+\ell_2^2\ell_3^2+\ell_3^2\ell_1^2)-\ell_1^4-\ell_2^4-\ell_3^4}
\end{align}
provides the area of a general triangle with side lengths $\ell_1$,
$\ell_2$, and $\ell_3$.
Combining Eqs.~(\ref{xi.ratios}) and (\ref{A.co.res}), and observing that
the area per lattice site $\Asite$ is twice the area of an elementary
lattice triangle, i.e.,
\begin{align}
\label{A.site}
\Asite=d_1h_1=2A_\text{tr}(d_1,d_2,d_3),
\end{align}
we obtain, asymptotically close to $\Tc$, the relation
\begin{align}
\label{Aco.res}
\Vco
&=
\f{2(1-z_2z_3)(z_2+z_3)}{(1-z_2^2)(1-z_3^2)}
\left(\f{{\xi_1}}{d_1}\right)^2\Asite,
\end{align}
which will be needed below.

\subsection{Free energy}
\label{free.energy}
In this section, we derive explicit results for the scaling behaviors of
$\fed{ex}$, $\fed{sf}$, and $\fed{fs}$ for periodic and antiperiodic
{\bc} and of $\fed{fs}$ for free {\bc}.
We provide scaling functions for their contributions per bulk correlation
volume, as well as standard scaling functions and critical Casimir
amplitudes for them.
The scaling functions and critical amplitudes will be expressed in terms
of the scaling variable $\xt$ from (\ref{xt}) and the anisotropy
parameters $r$ and $\th$.
For further reference we note that with (\ref{xt}) and
(\ref{nu.xi0p.xi0m}) follows
\begin{align}
\label{L.over.xiL}
L/\xi_L
&=
\begin{cases}
\ph{+}Lt/\xipz^{(L)}=\xt & \text{for }T>\Tc,\\[2ex]
-Lt/\ximz^{(L)}=-2\xt & \text{for }T<\Tc.
\end{cases}
\end{align}

For the two-dimensional Ising model, the leading singular behavior of $f$
in the bulk limit $L\to\infty$ behaves as $\propto t^2\ln|t|$.
The arbitrariness in splitting the constant under the logarithm between
singular and nonsingular contributions to $f$ prevents the required unique
splitting of $f$ according to (\ref{fns.fs}) and causes a violation of
scaling.
Consequently, $\Fcal$ does not exist.
On the other hand, such a splitting of $\fed{fs}$ and therefore the
function $\Gcal$ exist for all cases considered in this work. 
The existence of a corresponding splitting for $\fed{ex}$ and $\fed{sf}$,
and thus the existence of $\Acal$, depends on the {\bc}, as we will see
below.

It is useful to define the strip free energies per site $\fedh{}$ and per
bulk correlation volume $\fedt{}$ by
\bse
\label{f.site.co}
\begin{align}
\label{f.site}
\fedh{}(T,L)&=\Asite f(T,L),
\\
\label{f.co}
\fedt{}(T,L)&=\Vco f(T,L),
\end{align}
\ese
respectively.
Let analogous definitions hold for $\fedb$, $\fed{ex}$, $\fed{sf}$,
$\fed{fs}$, and, if a unique separation as in (\ref{fns.fs}) is defined,
also for their singular parts.

\subsubsection{Periodic and antiperiodic \bc}
\label{free.energy.pa}
The free energy per site defined in (\ref{f.site}) is, according to
Eq.~(71) of Ref.~\cite{LiHuLiWu99}, given by
\begin{align}
\label{f.3c}
\fedh{}(T&,L)
=
-C+\fedh{sf}(T,L)
\nn\\
&
-\f{1}{2\N}\sum_{j=0}^{\N-1}
\ln\f{f_1(\phi_j)+\sqrt{f_1^2(\phi_j)-f_2^2(\phi_j)}}{2},
\end{align}
with
\bse
\begin{align}
\phi_j&\equiv\f{2\pi(j+\f{1}{2})}{\N}, && \text{for periodic \bc},
\\
\phi_j&\equiv\f{2\pi j}{\N}, && \text{for antiperiodic \bc},
\end{align}
\ese
and where \cite{comment_LiHuLiWu99}
\bse
\label{f1.f2.X.Y.3c}
\begin{align}
f_1(\phi)&\equiv A_0-A_1\cos\phi,
\\
f_2(\phi)
&\equiv
\sqrt{(A_2+A_3)^2-4A_2A_3\sin^2(\phi/2)},
\end{align}
\ese
\bse
\label{Ai}
\begin{align}
A_0&\equiv(1+z_1^2)(1+z_2^2)(1+z_3^2)+8z_1z_2z_3,
\\
A_1&\equiv2z_1(1-z_2^2)(1-z_3^2),
\\
A_2&\equiv2z_2(1-z_3^2)(1-z_1^2),
\\
A_3&\equiv2z_3(1-z_1^2)(1-z_2^2),
\end{align}
\ese
and
\begin{align}
C&\equiv\ln\f{2}{\sqrt{(1-z_1^2)(1-z_2^2)(1-z_3^2)}},
\end{align}
with the qualification that the interface function $\fedh{sf}$ is missing
in Ref.~\cite{LiHuLiWu99}.
For $T>\Tc$ indeed $\fedh{sf}=0$.
For finite $L$, there is no phase transition and therefore the free energy
per site for $T<\Tc$ is obtained by analytic continuation of (\ref{f.3c}).
While for periodic \bc, this leaves $\fedh{sf}$ at zero, the $j=0$ term
has to be treated separately for antiperiodic {\bc} and we obtain
\bse
\label{f.if}
\begin{align}
\label{f.if.p}
\fedh{sf}^\pbc(T,L)
&=
0,
\\
\label{f.if.a}
\fedh{sf}^\abc(T,L)
&=
\f{\Th(-t)}{\N}\ln\f{(1-z_1)(1-z_2z_3)}{(1+z_1)(z_2+z_3)},
\end{align}
\ese
with the Heaviside step function $\Th$.
The interface contribution for antiperiodic {\bc} for $T<\Tc$ was
overlooked in Ref.~\cite{LiHuLiWu99}, leading to erroneous results there
for $T<\Tc$.
In particular, the $\theta<0$ plots for  $R=\infty$ in Fig.~3(a) and for
$L_y=\infty$ in Fig.~3(b) are missing a linearly rising part towards
$\th\to-\infty$.
This leads to incorrect statements between Eqs.~(78) and (79) of
Ref.~\cite{LiHuLiWu99} about the $\theta<0$ behavior of the infinitely
long cylinder.

Taking the limit $L\to\infty$ of (\ref{f.3c}), we obtain the bulk free
energy per site as
\begin{align}
\label{fb.3c}
\fedh{b}(T)=-C-\f{1}{4\pi}\int_0^{2\pi}d\phi
\ln\f{f_1(\phi)+\sqrt{f_1^2(\phi)-f_2^2(\phi)}}{2},
\end{align}
and we note that the critical point corresponds to
\begin{align}
\label{crit.point.1}
A_0-A_1-A_2-A_3=0,
\end{align}
which is equivalent to (\ref{crit.point.2}).

Combining (\ref{f.3c}) and (\ref{fb.3c}) according to
(\ref{fex.fsf.ffs.def}), we obtain in an obvious notation
\begin{align}
\fedh{fs}(T,L)
=&
\left[\f{1}{4\pi}\int_0^{2\pi}d\phi
-\f{1}{2\N}\sum_{j=0}^{\N-1}\right]
\times
\nn\\
&
\ln\f{f_1(\phi_{(j)})+\sqrt{f_1^2(\phi_{(j)})-f_2^2(\phi_{(j)})}}{2}.
\end{align}
From (\ref{f1.f2.X.Y.3c}) we obtain
\begin{align}
\label{f1.2.m.f2.2}
f_1^2&(\phi)-f_2^2(\phi)
=
\nn\\
&(A_0-A_1+A_2+A_3)(A_0-A_1-A_2-A_3)
\nn\\
&
+4\left[A_0A_1+A_2A_3-A_1^2\cos^2(\phi/2)\right]\sin^2(\phi/2).
\end{align}
Near $\Tc$ the quantity $A_0-A_1-A_2-A_3$ is small and an appropriate
approximation for $0\leq\phi\leq\pi$ may be written as
\begin{align}
\label{sq.ex.3c}
&\sqrt{f_1^2(\phi)-f_2^2(\phi)}
\approx
2\sqrt{A_0A_1+A_2A_3-A_1^2\cos^2\f{\phi}{2}}\,\sin\f{\phi}{2}
\nn\\
&{+}\f{\sqrt{(A_0-A_1)A_1+A_2A_3}}{\N}
\left(\sqrt{\xt^2+\N^2\phi^2}-\N\phi\right),
\end{align}
where we have used that through linear order in $t$
\begin{align}
\pm\N
&\sqrt{\f{(A_0-A_1+A_2+A_3)(A_0-A_1-A_2-A_3)}{(A_0-A_1)A_1+A_2A_3}}
\nn\\
&=
Lt/\xipz^{(1)}=\xt
\quad\text{for $T\gtrless\Tc$},
\end{align}
which follows from (\ref{crit.point.3}), (\ref{z1.z1c.exp}),
(\ref{L.over.xiL}), and (\ref{Ai}).
An approximation analogous to (\ref{sq.ex.3c}) holds for
$\pi\leq\phi\leq2\pi$, which is found by replacing $\phi\to2\pi-\phi$
there, under which (\ref{f1.2.m.f2.2}) is invariant.
Thus we may write
\begin{widetext}
\begin{align}
\label{fex.pa.int.sum.tr}
\fedh{fs}(T,L)
&\approx
\left[\f{1}{4\pi}\int_0^{2\pi}d\phi
-\f{1}{2\N}\sum_{j=0}^{\N-1}\right]
\ln\f{A_0-A_1\cos\phi_{(j)}
+2\sqrt{A_0A_1+A_2A_3-A_1^2\cos^2\f{\phi_{(j)}}{2}}\,\sin\f{\phi_{(j)}}{2}}{2}
\nn\\
&\ph{=}
+\f{\sqrt{(A_0-A_1)A_1+A_2A_3}}{(A_0-A_1)\N}\left[
\f{1}{2\pi}\int_0^{2\pi}d\phi-\f{1}{\N}\sum_{j=0}^{\N-1}\right]
\left(\sqrt{\xt^2+\N^2\phi_{(j)}^2}-\N\phi_{(j)}\right).
\end{align}
\end{widetext}
Using the results (\ref{Iabm}) and (\ref{delta.sq.p}) for periodic {\bc}
and (\ref{Iabl}) and (\ref{delta.sq.a}) for antiperiodic \bc, we obtain
\begin{align}
\label{fex.pa.raw.tr}
\N^2\fedh{fs}^\psabc(T,L)
&\approx
\f{\sqrt{(A_0-A_1)A_1+A_2A_3}}{(A_0-A_1)}\Ipsm(\xt),
\end{align}
with
\begin{align}
\label{Ipm}
\Ipm(x)
&\equiv
-\f{1}{\pi}\int_0^\infty d\om\ln\left(1\pm e^{-\sqrt{x^2+\om^2}}\right),
\end{align}
and where we note that
\begin{align}
\label{Ipm.0}
\Ip(0)
&=
-\pi/12,
&
\Im(0)
&=
\pi/6.
\end{align}

Close to the critical point, we obtain, after some algebra and with the
help of (\ref{Aco.res}),
\begin{align}
\label{de.prefac.1}
\f{\sqrt{(A_0-A_1)A_1+A_2A_3}}{(A_0-A_1)}
\approx\left(\f{{\xi_1}}{d_1}\right)^2\f{\Asite}{\Vco}.
\end{align}
Combining this with (\ref{fex.pa.raw.tr}) gives, close to $\Tc$,
\begin{align}
\label{L2.fhfs}
\fedh{fs}^\psabc(T,L)
&\approx
\f{\xi_1^2}{L^2}\f{\Asite}{\Vco}\Ipsm(\xt).
\end{align}
Using (\ref{xi1.xiL}), (\ref{z1.z1c.exp}), (\ref{Aco.res}), and
(\ref{L.over.xiL}), we obtain for the interface contribution
(\ref{f.if.a}) to the free energy per site for antiperiodic {\bc} for
temperatures close to $\Tc$
\begin{align}
\fedh{sf}^\abc(T,L)
&\approx
-\f{\xi_1^2}{L^2}\f{\Asite}{\Vco}\Th(-\xt)\xt.
\end{align}

For periodic and antiperiodic \bc, the free energy contributions
$\fed{ex}$, $\fed{sf}$, and $\fed{fs}$ are dominated by their leading
singular parts $\fed{ex,s}$, $\fed{sf,s}$, and $\fed{fs,s}$, respectively.
With (\ref{xi1.xiL}), (\ref{L.over.xiL}), and (\ref{f.site.co}), we obtain
the asymptotic singular finite-size and interface parts of the free energy
per bulk correlation volume as
\begin{align}
\label{ffsco}
\fedt{fs,s}^\psabc(t,L)
&=
\begin{cases}
\Ipsm(\xt)/\xt^2 & \text{for }T>\Tc,\\[2ex]
\Ipsm(\xt)/(4\xt^2) & \text{for }T<\Tc,
\end{cases}
\end{align}
and
\bse
\label{fsfco}
\begin{align}
\fedt{sf,s}^\pbc(t,L)
&=
0,
\\
\fedt{sf,s}^\abc(t,L)
&=
-\Th(-\xt)/(4\xt),
\end{align}
\ese
respectively.
With (\ref{fsf.ffs.def}) follows the asymptotic singular excess free
energy per correlation volume
\bse
\label{fexco}
\begin{align}
\fedt{ex,s}^\pbc(t,L)
&=
\fedt{fs,s}^\pbc(t,L),
\\[1ex]
\fedt{ex,s}^\abc(t,L)
&=
\begin{cases}
\fedt{fs,s}^\abc(t,L) & \text{for }T>\Tc,\\[2ex]
[\Im(\xt)-\xt]/(4\xt^2) & \text{for }T<\Tc.
\end{cases}
\end{align}
\ese
Eqs.~(\ref{ffsco})--(\ref{fexco}) represent central results of the current
section.
Their most remarkable feature is that their right hand sides depend only
on the scaling variable $\xt$ and not on the parameters $\ar$ and $\th$
describing the anisotropy of the system at large distances (nor on any
other details of how the anisotropy is realized).
We conjecture that the right hand sides of (\ref{ffsco})--(\ref{fexco})
represent universal scaling functions in the bulk universality class of
the two-dimensional Ising model.
In Sec.~\ref{d-dim.film.pa.bc}, we will explain their independence of $\ar$
and $\th$ in the more general context of $d$-dimensional films.

Next we derive standard free energy scaling functions and critical
amplitudes.
Combining (\ref{A.co.def}) and (\ref{xi.L.2}) gives
\begin{align}
\label{xi1sq.over.Aco}
\Vco/\xi_L^2
&=
\ar\cos^2\th+\ar^{-1}\sin^2\th,
\end{align}
so that multiplying (\ref{ffsco}) and (\ref{fsfco}) by $L^2/\Vco$,
while observing (\ref{L.over.xiL}), gives asymptotically
\begin{align}
\label{Gcal.pa.3c}
L^2\fed{fs,s}^\psabc(t,L)
&=
\Gcal^\psabc(\xt,\ar,\th)
\nn\\
&=
\f{\Ipsm(\xt)}{\ar\cos^2\th+\ar^{-1}\sin^2\th},
\end{align}
and
\begin{align}
L^2\fed{sf,s}^\abc(t,L)
&=
-\f{\Th(-\xt)\xt}{\ar\cos^2\th+\ar^{-1}\sin^2\th},
\end{align}
respectively.
Therefore, we have
\begin{align}
\label{Acal.pa.3c}
L^2&\fed{ex,s}^\psabc(t,L)
=\Acal^\psabc(\xt,\ar,\th)
\nn\\
&=
\begin{cases}
\Gcal^\pbc(\xt,\ar,\th) & \text{for periodic \bc,}\\[2ex]
\ds\f{\Im(\xt)-\Th(-\xt)\xt}{\ar\cos^2\th+\ar^{-1}\sin^2\th}
 & \text{for antiperiodic \bc,}
\end{cases}
\end{align}
where $\Acal^\psabc(\xt,\ar,\th)$ and $\Gcal^\psabc(\xt,\ar,\th)$ are the
scaling functions of the excess free energy density and the finite-size
contribution to the free energy density, respectively.
For the two-dimensional Ising model, they generalize the isotopic-case
structure (\ref{Acal.Gcal}) to the anisotropic case.
We conjecture that $\Acal^\psabc$ and $\Gcal^\psabc$ are universal
functions of $\xt$, $\ar$, and $\th$ for all infinite-strip systems in the
bulk universality class of the two-dimensional Ising model with the
appropriate {\bc} in the $L$ direction.
Note that the scaling functions depend on the anisotropy parameters $\ar$
and $\th$ only through a geometric factor.

\begin{figure}
\begin{center}
\includegraphics[width=8.5cm,angle=0]{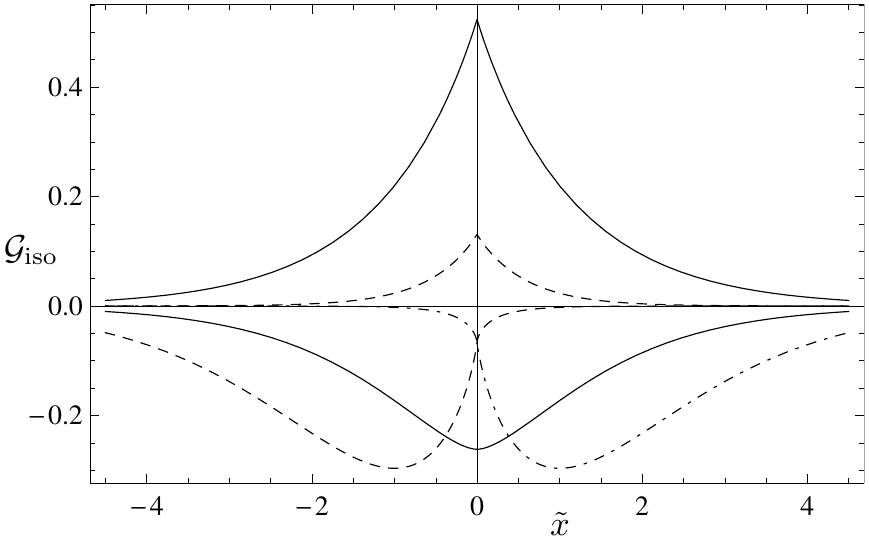}
\end{center}
\caption{\label{Gisofig}
Scaling functions $\Gcaliso^\pbc(\xt)$ (solid, $<0$), $\Gcaliso^\abc(\xt)$
(solid, $>0$), $\Gcaliso^\zzbc(\xt)$ (dashed, $<0$), $\Gcaliso^\ppbc(\xt)$
(dot-dashed, $<0$), and $\Gcaliso^\zpbc(\xt)$ (dashed, $>0$).}
\end{figure}

For the isotropic case $\ar=1$, the results (\ref{Gcal.pa.3c}) and
(\ref{Acal.pa.3c}) reduce to
\begin{align}
\label{Gcal.iso.pa}
\Gcaliso^\psabc(\xt)
&=
\Ipsm(\xt)
\end{align}
and
\begin{align}
\label{Acal.iso.pa}
\Acal_\text{iso}^\psabc(\xt)=
\begin{cases}
\Gcaliso^\pbc(\xt) & \text{for periodic \bc,}\\[2ex]
\ds\Im(\xt)-\Th(-\xt)\xt  & \text{for antiperiodic \bc,}
\end{cases}
\end{align}
and thus we may relate
\bse
\label{Gcal.pa}
\begin{align}
\!\!\!\Gcal^\psabc(\xt,r,\th)
&=
\left(\ar\cos^2\th{+}\ar^{-1}\sin^2\th\right)^{-1}
\Gcaliso^\psabc(\xt),
\\
\!\!\!\Acal^\psabc(\xt,r,\th)
&=
\left(\ar\cos^2\th{+}\ar^{-1}\sin^2\th\right)^{-1}
\Acaliso^\psabc(\xt).
\end{align}
\ese
With $x_\perp=\xt$, our result for $\Gcaliso^\pbc(\xt)$ is identical to
$\Th_\perp(x_\perp,0)$ from Eq.~(57) of Ref.~\cite{HuGrSc10}.
Even for the isotropic limit, $\Gcal^\abc$ and $\Acal^\abc$ appear not
to have been previously published.
$\Gcaliso^\pbc(\xt)$ and $\Gcaliso^\abc(\xt)$ are shown in
Fig.~\ref{Gisofig}.
A cross check of $\Gcaliso^\psabc$ and $\Acal_\text{iso}^\psabc$
with published scaling functions for the Casimir force is provided in
Appendix~\ref{iso.scal.func.pa}.

With (\ref{Ipm.0}), the critical Casimir amplitudes
$\De(\ar,\th)=\Gcal(0,\ar,\th)$ corresponding to (\ref{Gcal.pa.3c}) are
\bse
\label{De.p.a}
\begin{align}
\De^\pbc(\ar,\th)
&=\f{-\pi/12}{\ar\cos^2\th+\ar^{-1}\sin^2\th},
\\
\De^\abc(\ar,\th)
&=\f{\pi/6}{\ar\cos^2\th+\ar^{-1}\sin^2\th},
\end{align}
\ese
which we expect to be universal functions of $\ar$ and $\th$ within the
bulk universality class of the two-dimensional Ising model.

\subsubsection{Free \bc}
\label{free.energy.00}
The free energy per site defined in (\ref{f.site}) is, according to
Eq.~(A15) of Ref.~\cite{InNiWa86}, for free {\bc} given by
\cite{comment_InNiWa86_b}
\begin{align}
\label{f.3c.rect.fixed}
&\fedh{rect}^\zzbc(T,\Lp,L)
=
-\ln(2\cosh K_1\cosh K_2\cosh K_3)
\nn\\
&
+\f{1}{\N}\ln(\cosh K_2\cosh K_3)
-\f{1}{\Np}\sum_\vt\ln|1{-}z_1e^{i\vt}|
\nn\\
&
-\f{1}{2\Np\N}
\sum_\vt\ln(p_+\la_+^{\N-1}+p_-\la_-^{\N-1}),
\end{align}
with
\begin{align}
\vt&=\f{(2p-1)\pi}{\Np}, & p&=1,\ldots,\Np,
\end{align}
\bse
\label{p.la.pm}
\begin{align}
\label{ppm}
p_\pm&=\pm\f{A-\la_\mp-\bar{a}|E|}{\sqrt{(A-F)^2+4|EC|}},
\\
\label{lapm}
\la_\pm
&=
\f{A+F\pm\sqrt{(A-F)^2+4|EC|}}{2},
\end{align}
\ese
and
\bse
\begin{align}
\bar{a}&=-2z_1|\sin\vt|/|1-z_1e^{i\vt}|^2,
\\
b&=(1-z_1^2)/|1-z_1e^{i\vt}|^2,
\\
A&=(\al^2+\eta^2+2\al\eta\cos\vt)/|1-z_1e^{i\vt}|^2,
\\
E&=(2\be\ga\sin\vt)/|1-z_1e^{i\vt}|^2,
\\
C&=(2\al\eta\sin\vt)/|1-z_1e^{i\vt}|^2,
\\
F&=(\be^2+\ga^2+2\be\ga\cos\vt)/|1-z_1e^{i\vt}|^2,
\\
\al&=z_1+z_2z_3,
\\
\be&=z_2+z_3z_1,
\\
\ga&=z_3+z_1z_2,
\\
\eta&=-(1+z_1z_2z_3).
\end{align}
\ese
The strip free energy per site is obtained as
\begin{align}
\label{f.3c.strip.fixed}
&\fedh{}^\zzbc(T,L)
=
\lim_{\Lp\to\infty}\fedh{rect}^\zzbc(T,\Lp,L)
\nn\\
&=
-\ln(2\cosh K_1\cosh K_2\cosh K_3)
\nn\\
&\ph{=}
+\f{1}{\N}\ln(\cosh K_2\cosh K_3)
-\f{1}{2\pi}\int_0^{2\pi}d\vt\ln|1-z_1e^{i\vt}|
\nn\\
&\ph{=}
-\f{1}{4\pi\N}\int_0^{2\pi}d\vt\ln[
(p_+\la_+^{\N-1}+p_-\la_-^{\N-1})].
\end{align}
Taking the limit $L\to\infty$ and observing that $\la_+>\la_-$, we obtain
the bulk free energy per site (\ref{fb.3c}) in the alternative
representation
\begin{align}
\label{f.3c.bulk}
&\!\!\fedh{b}(T)
=
\lim_{L\to\infty}\fedh{}^\zzbc(T,L)
\nn\\
&=
-\ln(2\cosh K_1\cosh K_2\cosh K_3)
\nn\\
&\ph{=}
-\f{1}{2\pi}\int_0^{2\pi}d\vt\ln|1-z_1e^{i\vt}|
-\f{1}{4\pi}\int_0^{2\pi}d\vt\ln \la_+.
\end{align}
Thus the excess free energy per site is
\begin{align}
\label{fex.3c.strip.fixed}
\fedh{ex}^\zzbc(T,L)
&=
\fedh{sf}^\zzbc(T,L)+\fedh{fs}^\zzbc(T,L),
\end{align}
with
\begin{align}
\label{fif.3c.strip.fixed}
\fedh{sf}^\zzbc(T,L)
&=
\f{1}{\N}\ln(\cosh K_2\cosh K_3)
\nn\\
&\ph{=}
-\f{1}{4\pi\N}\int_0^{2\pi}d\vt\ln\f{p_+}{\la_+}
\end{align}
and
\begin{align}
\label{ffs.3c.strip.fixed}
&\fedh{fs}^\zzbc(T,L)
=
\nn\\
&\ph{=}
-\f{1}{4\pi\N}\int_0^{2\pi}d\vt\ln\left[
1+\f{p_-}{p_+}\left(\f{\la_-}{\la_+}\right)^{\N-1}\right],
\end{align}
where we already anticipate that the latter expression does not contain
any surface or interface terms.
Near $\Tc$, dominant contributions to the integral in
(\ref{ffs.3c.strip.fixed}) arise only near $\vt=0$ and $\vt=2\pi$.
Since the integrand in (\ref{ffs.3c.strip.fixed}) is invariant under
$\vt\to2\pi-\vt$, it is sufficient to consider small positive $\vt$.
Employing the small-$t$ expansion (\ref{z1.z1c.exp}), we obtain
\bse
\begin{align}
A+F&\approx2(1-z_2z_3)^2,
\\
\sqrt{(A-F)^2+4|EC|}
&\approx
\f{2(1-z_2z_3)^2}{\N}\sqrt{\xbar^2+\om^2},
\\
A-F-2\bar{a}|E|
&\approx
\f{2(1-z_2z_3)^2}{\N}\xbar,
\end{align}
\ese
with
\begin{align}
\label{om}
\om
&\equiv
\f{(1-z_2^2)(1-z_3^2)}{2(1-z_2z_3)(z_2+z_3)}\N\vt
\end{align}
and
\begin{align}
\label{xbar.Ising}
\xbar
&\equiv
\f{(1-z_2^2)(1-z_3^2)}{2(1{-}z_2z_3)(z_2{+}z_3)}\f{\N d_1t}{\xipz^{(1)}}
\approx
\f{\xi_1^2}{\Vco}\f{Lt}{\xipz^{(1)}}
\approx
\f{\xi_\infty\xi_L}{\xig\xik}\f{Lt}{\xipz^{(L)}}
\nn\\
&\approx
\left[(r\sin^2\th+r^{-1}\cos^2\th)
(r\cos^2\th+r^{-1}\sin^2\th)\right]^{-1/2}\xt
\nn\\
&=
\left[1+\fr{1}{4}(r-r^{-1})^2\sin^2(2\th)\right]^{-1/2}\xt,
\end{align}
where (\ref{xt}), (\ref{L23}), (\ref{L.L23}),
(\ref{r}), (\ref{xi.infty.L.2}), (\ref{xi1.xiinfty}),
(\ref{nu.xi0p.xi0m}), (\ref{A.site}), and (\ref{Aco.res}) have been used.
It follows that
\bse
\begin{align}
\la_\pm
&\approx
(1-z_2z_3)^2\left(1\pm\f{\sqrt{\xbar^2+\om^2}}{\N}\right),
\\
p_\pm
&\approx
\f{\sqrt{\xbar^2+\om^2}\pm\xbar}{2\sqrt{\xbar^2+\om^2}},
\end{align}
\ese
and therefore that
\bse
\label{pmopp.lmolp}
\begin{align}
\label{ppolp}
\f{p_+}{\la_+}
&\approx
\f{1}{2(1-z_2z_3)^2}\left(1+\f{\xbar}{\sqrt{\xbar^2+\om^2}}\right),
\\
\f{p_-}{p_+}
&\approx
\f{\sqrt{\xbar^2+\om^2}-\xbar}{\sqrt{\xbar^2+\om^2}+\xbar},
\\
\left(\f{\la_-}{\la_+}\right)^{\N-1}
&\approx
e^{-2\sqrt{\xbar^2+\om^2}}.
\end{align}
\ese
From the expression for $p_+/\la_+$ we conclude that the second term on
the right hand side of (\ref{fif.3c.strip.fixed}) receives a nonsingular
contribution $N^{-1}[\ln(1-z_2z_3)-\f{1}{2}\ln2]$ plus a contribution to
the integral that is not restricted to small $\vt$ or $2\pi-\vt$, since $\ln(1+\xbar/\sqrt{\xbar^2+\om^2})\approx\xbar/\om$ for large $\om$.
This points towards a singular contribution that logarithmically violates
scaling of $\fed{sf}^\zzbc$, and we will not consider this quantity in
what follows.

Combining (\ref{ffs.3c.strip.fixed}), (\ref{om}), and (\ref{pmopp.lmolp}),
we obtain in the large-$\N$ limit close to $\Tc$ for the finite-size term 
\begin{align}
\label{fhfs.3c.strip.fixed.1}
\N^2\fedh{fs}^\zzbc(T,L)
&\approx
\f{2(1-z_2z_3)(z_2+z_3)}{(1-z_2^2)(1-z_3^2)}\If(\xbar),
\end{align}
with
\begin{align}
\label{If}
&\If(x)
\nn\\
&\equiv
-\f{1}{2\pi}\int_0^\infty d\om
\ln\left(1+\f{\scs\sqrt{x^2+\om^2}-x}{\scs\sqrt{x^2+\om^2}+\xbar}
e^{-2\sqrt{x^2+\om^2}}\right).
\end{align}
We note that
\begin{align}
\label{If.0}
\If(0)&=\Ip(0)/4=-\pi/48,
\end{align}
which results from comparing (\ref{If.0}) with (\ref{Ipm}) and
(\ref{Ipm.0}).

Combining (\ref{L23}), (\ref{L.L23}), (\ref{A.co.def}),
(\ref{xi1.xiinfty}), (\ref{A.site}), (\ref{Aco.res}), and
(\ref{fhfs.3c.strip.fixed.1}), we obtain, close to $\Tc$,
\begin{align}
\fedh{fs}^\zzbc&(T,L)
\approx
\f{\xig^2\xik^2}{L^2\xi_\infty^2}\f{\Asite}{\Vco}\If(\xbar).
\end{align}
For free \bc, the free energy contribution $\fed{fs}$ is dominated by its
leading singular part $\fed{fs,s}$.
With (\ref{xi.crit}), (\ref{nu.xi0p.xi0m}), (\ref{f.site.co}), and
(\ref{xbar.Ising}), we obtain for the asymptotic singular finite-size part
of the free energy per bulk correlation volume
\begin{align}
\label{ffsco.00bc}
\fedt{fs,s}^\zzbc(t,L)
&=
\begin{cases}
\If(\xbar)/\xbar^2 & \text{for }T>\Tc,\\[2ex]
\If(\xbar)/(4\xbar^2) & \text{for }T<\Tc,
\end{cases}
\end{align}
where Eq.\ (\ref{xbar.Ising}) gives $\xbar$ as a function of $\xt$, $r$,
and $\th$.
Eq.~(\ref{ffsco.00bc}) represents a central result of the current
section.
Its most remarkable feature is that its right hand side depends only on
the single variable $\xbar$.
We conjecture that the right hand side of (\ref{ffsco.00bc}) represents a
universal scaling function in the bulk universality class of the
two-dimensional Ising model.
In Sec.~\ref{d-dim.film.fixed.bc}, we will explain its independence of
$\ar$ and $\th$ in the more general context of $d$-dimensional films.

Next we derive standard free energy scaling functions and critical
amplitudes.
Combining (\ref{A.co.def}) and (\ref{xi.infty.2}) gives, in the asymptotic
critical domain, 
\begin{align}
\label{xi1sq.over.Aco.fixed}
\Vco/\xi_\infty^2
&=
\ar\sin^2\th+\ar^{-1}\cos^2\th,
\end{align}
so that, multiplying (\ref{ffsco.00bc}) by $L^2/\Vco$ while observing
(\ref{xi.crit}), (\ref{nu.xi0p.xi0m}), (\ref{xbar.Ising}), and
(\ref{xi1sq.over.Aco.fixed}), gives asymptotically
\begin{align}
\label{Gcal.00}
L^2\fed{fs,s}^\zzbc(t,L)
&=
\Gcal^\zzbc(\xt,r,\th)
\nn\\
&=
(\ar\sin^2\th+\ar^{-1}\cos^2\th)\If(\xbar),
\end{align}
with $\xbar$ from (\ref{xbar.Ising}).
We conjecture that $\Gcal^\zzbc$ is a universal function of $\xt$, $\ar$,
and $\th$ in the bulk universality class of the two-dimensional Ising
model.
Note that the scaling function $\Gcal^\zzbc$ depends on the anisotropy
parameters $\ar$ and $\th$ only through the geometric factors on the right
hand sides of (\ref{xbar.Ising}) and (\ref{Gcal.00}).

With (\ref{If.0}), the related Casimir amplitude is
\begin{align}
\label{De00}
\De^\zzbc(r,\th)
&=\Gcal^\zzbc(0,r,\th)
\nn\\
&=
-\f{\pi}{48}(\ar\sin^2\th+\ar^{-1}\cos^2\th).
\end{align}
For the isotropic case, Eq.\ (\ref{Gcal.00}) reduces to
\begin{align}
\label{G00.iso}
\Gcaliso^\zzbc(\xt)
&=
\If(\xt),
\end{align}
and thus we may relate
\begin{align}
\label{G00}
\Gcal^\zzbc(\xt,r,\th)
&=
\left(\ar\sin^2\th+\ar^{-1}\cos^2\th\right)
\Gcaliso^\zzbc(\xbar),
\end{align}
with $\xbar$ from (\ref{xbar.Ising}).
With $x=2\xt$, our result for $\Gcaliso^\zzbc(\xt)$ is identical to
$X_\text{ex}^{(o,o)}(x)$ for ordinary {\bc} provided at the beginning of
Sec.~12.1.2 in \cite{BrDaTo00} without derivation.
$\Gcaliso^\zzbc(\xt)$ is shown in Fig.~\ref{Gisofig}.
A derivation of $\Gcaliso^\zzbc$ from published scaling functions
for the Casimir force is provided in Appendix~\ref{iso.scal.func.ff}.

As another cross check, we compare our results with the Casimir amplitude
found in Ref.~\cite{InNiWa86}, where it is defined by
\begin{align}
\label{De.INW}
\De_\text{INW}\equiv N^2\fedh{fs,s}(0,L),
\end{align}
which, observing (\ref{L23}), (\ref{L.L23}), and (\ref{A.site}), is
related to $\De(\ar,\th)=\Gcal(0,\ar,\th)$ by
\begin{align}
\label{DeINW.De}
\De_\text{INW}=(d_1/h_1)\De(r,\th).
\end{align}
Ref.~\cite{InNiWa86} provides the result
\begin{align}
\label{De.INW.00}
\De_\text{INW}^\zzbc
&=-(\pi/48)\eta,
\end{align}
with
\begin{align}
\label{eta}
\eta
&\equiv
\f{C_2^2}{S_1+S_2}
=\f{d_1}{h_1}(r\sin^2\th+r^{-1}\cos^2\th),
\end{align}
where $C_i\equiv\cosh(2K_i)$ and $S_i\equiv\sinh(2K_i)$.
The second equality in (\ref{eta}) is obtained by using 
(\ref{A.co.def}), (\ref{r}), (\ref{xi.L.2}), (\ref{xi1.xiinfty}),
(\ref{zi}), (\ref{A.site}), and (\ref{Aco.res}), as well as  the
criticality condition
\begin{align}
S_1S_2+S_2S_3+S_3S_1=1,
\end{align}
which is equivalent to (\ref{crit.point.2}) and (\ref{crit.point.1}).
Comparison of (\ref{De00}) and (\ref{DeINW.De})--(\ref{eta}) shows that 
our result for $\De^\zzbc(r,\th)$ agrees with the corresponding result in
Ref.~\cite{InNiWa86}.
However, while $\De^\zzbc(r,\th)$ is universal, this is not the case
for $\De_\text{INW}$ as defined in (\ref{De.INW}), since $\De_\text{INW}$
depends on lattice details as manifested by the factor $d_1/h_1$ appearing
in (\ref{DeINW.De}) and (\ref{eta}).

\section{Film in $d$ dimensions}
\label{d-dim.film}
The results of Sec.\ \ref{I2.fe} may be analyzed in the more general
context of a $d$-dimensional film of thickness $L$, i.e., for a system in
an $L\times\infty^{d-1}$ geometry near a $d$-dimensional bulk critical
point.
Note that, within this section, we refer without further mentioning always
to the asymptotic critical domain.

\subsection{Bulk correlation lengths}
\label{corr.len.film}
Weakly anisotropic systems may be related to corresponding isotropic
systems by an anisotropic scale transformation, see, e.g., \cite{Ca87}.
Such a transformation may be realized as a shear transformation
(see \cite{ChDo04,Do08,DiCh09} and \cite{Do06} for the use of such a
transformation in the context of critical phenomena for field-theoretic
models and for $\vp^4$ lattice models, respectively)
\begin{align}
\xb'=\Mb\xb,
\end{align}
where $\xb$ and $\xb'$ represent the $d$ Cartesian coordinates of position
vectors in the original anisotropic system and the related isotropic
system, respectively, and where the $d\times d$ matrix $\Mb$ may be
decomposed according to
\begin{align}
\label{M.SR}
\Mb=\Sb\Rb
\end{align}
into a rotation, represented by an orthogonal matrix $\Rb$, and a
subsequent rescaling, represented by a real diagonal matrix $\Sb$.

It is then straightforward to adapt the above transformation to the case
treated here, where the long-distance correlations of the anisotropic bulk
system are described by a correlation lengths ellipsoid represented by a
tensor $\Xib$ that is diagonalized by $\Rb$, so that
\begin{align}
\Rb\Xib\Rb^{-1}=\diag(\xi_1^2,\xi_2^2,\ldots,\xi_d^2),
\end{align}
with the correlation lengths $\xi_1,\ldots,\xi_d$ along the principal
axes of the ellipsoid, and where $\Sb$ is the volume-conserving
($\det\Sb=1$) rescaling matrix
\begin{align}
\Sb=\Vco^{1/d}\diag(\xi_1^{-1},\xi_2^{-1},\ldots,\xi_d^{-1}),
\end{align}
with the correlation volume defined by
\begin{align}
\Vco\equiv\xi_1\xi_2\cdots\xi_d=(\det\Xib)^{1/2}.
\end{align}
Thus the correlation length in the isotropic system is
\begin{align}
\label{Vco}
\xi=\Vco^{1/d}.
\end{align}
Consequently, the squared correlation lengths tensor in the primed system
is
\begin{align}
\label{Xibp.Xib}
\Xib'=\Mb\Xib\Mb^T=\Vco^{2/d}\unit,
\end{align}
with the $d\times d$ unit matrix $\unit$.
Note that
\begin{align}
\label{Xib.inv}
\Xib^{-1}=\Vco^{-2/d}\Mb^T\Mb.
\end{align}

In the $\vp^4$ lattice model and in the field-theoretic contexts of
Refs.~\cite{Do06} and \cite{ChDo04,Do08}, respectively, the shear
transformation is written as $\lab^{-1/2}\Ub$, where $\Ub$ corresponds to
our rotation $\Rb$ and $\lab^{-1/2}$ corresponds to $\Sb$ with the
qualification that, in contrast to our definition, $\lab$ is not
necessarily volume-conserving.
In Refs.~\cite{ChDo04,Do06,Do08}, a matrix $\Ab=\Ub^{-1}\lab\Ub$ was defined,
so that $\Abar\equiv\Ab/(\det\Ab)^{1/d}$ may be used to parameterize the
long-distance anisotropy \cite{Do08}.
Since we choose $\Sb$ to be volume-conserving, our conventions imply
\begin{align}
\label{Ab}
\Ab=\Abar=(\Mb^T\Mb)^{-1}=(\det\Xib)^{-1/d}\Xib,
\end{align}
with
\begin{align}
\label{det.A}
\det\Ab=\det\Abar=1.
\end{align}

Note that $\Abar$ is defined here through the physical correlation lengths
in the asymptotic critical domain.
In contrast, explicit versions of $\Abar$ were obtained in
\cite{ChDo04,Do06,Do08} by requiring the related shear transformation to
lead to a transformed Hamiltonian, whose expansion in small wave numbers
$k$ is isotropic through order $k^2$.
While for standard $\vp^4$ field theory these definitions should coincide
due to an exact mapping between the anisotropic and isotropic bulk
Hamiltonians \cite{ChDo04}, there is no reason to believe that this
procedure generally leads also for lattice models to $\Abar$ as defined
here.
We will return to this issue in Sec.~\ref{summary}.

Let $\nhb$ be a unit vector that is orthogonal to the film boundaries 
and define the vector
\begin{align}
\label{Lb}
\Lb=L\nhb.
\end{align}
For two-dimensional systems such as the Ising model
treated in Sec.~\ref{I2.fe}, it is convenient to choose coordinates, where
\begin{align}
\label{nhat}
\nhb^T=(1,0),
\end{align}
so that, with $r$ and $\th$ as defined in Sec.\ \ref{corr.lengths.3c}, we
have
\begin{align}
\Rb=
\begin{pmatrix} 
\cos\th & \sin\th \\
-\sin\th & \cos\th
\end{pmatrix},
\end{align}
and
\begin{align}
\Sb=
\begin{pmatrix} 
r^{1/2} & 0 \\
0 & r^{-1/2}
\end{pmatrix},
\end{align}
from which follows with (\ref{M.SR}) and (\ref{Ab})
\begin{align}
\label{Abar.inv}
\Abar^{-1}=
\begin{pmatrix}
r\cos^2\th+r^{-1}\sin^2\th & \fr{1}{2}(r{-}r^{-1})\sin(2\th) \\
\fr{1}{2}(r{-}r^{-1})\sin(2\th) & r\sin^2\th+r^{-1}\cos^2\th
\end{pmatrix}.
\end{align}
Therefore, the parameterization of the anisotropy of a $d$-dimensional
system by $\Abar$ and the corresponding film orientation by $\nhb$
reduces, for a two-dimensional system, naturally to the parameterization
by $\ar$ and $\th$, as manifested, for the two-dimensional Ising model, by
Eqs.\ (\ref{Gcal.pa.3c})--(\ref{Acal.pa.3c}), (\ref{De.p.a}),
(\ref{Gcal.00}), and (\ref{De00}).

\subsection{Free energy}
For a $d$-dimensional system with restricted geometry, the system shape
and the {\bc} are generally transformed in a nontrivial way by the shear
transformation.
A major simplification arises for films with \mbox{(anti-)}periodic, fixed
or free {\bc}, since (i) a film is transformed into another  film and (ii)
the {\bc} are invariant.
This means that we can express film free energy scaling functions of the
anisotropic system in terms of the corresponding scaling functions of the
isotropic system with the same geometry and {\bc}.
Then the modifications for the anisotropic system can be represented by
geometric factors, as we already explicitly observed in Sec.~\ref{I2.fe}
for the two-dimensional Ising model and as will be detailed for
$d$-dimensional film systems below.

For notational simplicity, we formulate what follows for the singular
part $\fed{s}$ of the free energy density $f$, even though for the
particular case of the two-dimensional Ising model there is no unique
separation of regular part $\fed{s}$ and singular part $\fed{ns}$ and
consequently $\Fcal$ does not exist, in contrast to $\Acal$ and/or
$\Gcal$, depending on the \bc.
If universality continues to hold for anisotropic $d$-dimensional film
systems, we expect Eq.\ (\ref{Fcal}) to be replaced by
\begin{align}
\label{Fcal.aniso}
L^df_s(t,L)=\Fcal(\xt,\Abar,\nhb),
\end{align}
where $\Fcal$ is a universal function of its arguments and the imposed
\bc.
Eqs.\ (\ref{Acal.Gcal}) are replaced analogously.

Due to the different ways a film transforms for periodic and antiperiodic
{\bc} on the one hand and for fixed and free {\bc} on the other hand, we
treat these cases separately.
\begin{figure}
\begin{center}
\includegraphics[width=8.5cm,angle=0]{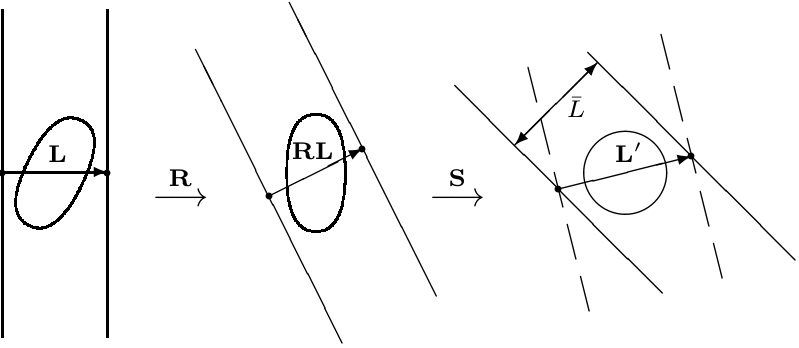}
\end{center}
\caption{\label{shear.trafo}
Illustration of the shear transformation from an anisotropic to an
isotropic film for $d=2$, i.e., for an infinite strip.
Shown are the behavior of the strip boundaries, the correlation ellipse,
and the vector $\Lb$ under a rotation $\Rb$ and a subsequent rescaling
$\Sb$ along the horizontal and vertical coordinate axes.
$\Lbar$ is the thickness of the resulting strip, needed for free and fixed
\bc.
For periodic and antiperiodic {\bc}, the indicated dots at opposite sides
of the original, rotated, and rescaled strips, connected by the vectors
$\Lb$, $\Rb\Lb$, and $\Lb'=\Sb\Rb\Lb$, respectively, are physically
identical.
For these {\bc}, translational invariance allows to define new strip
boundaries (dashed) that are perpendicular to the direction of
(anti-)periodicity, so that $L'=|\Lb'|$ is the thickness of the resulting
strip.}
\end{figure}

\subsubsection{Periodic and antiperiodic \bc}
\label{d-dim.film.pa.bc}
Consider a film with periodic or antiperiodic {\bc} and a length $L$ of
\mbox{(anti-})periodicity in the direction represented by the unit vector
$\nhb$.
The shear transformation $\Mb$ transforms the film into another film,
albeit with a different vector
\begin{align}
\label{Lbp}
\Lb'=\Mb\Lb
\end{align}
describing the direction and length \mbox{$L'=|\Lb'|$} of
\mbox{(anti-})periodicity of the isotropic system.
For \mbox{(anti-})periodic {\bc}, the orientation of the film boundaries
is not unique.
We choose them such that $\Lb$ is orthogonal to the boundaries of the
original film and $\Lb'$ is orthogonal to the boundaries of the film with
isotropic bulk correlation lengths, see Fig.~\ref{shear.trafo}.
Other choices lead to skewed {\bc}.
Results for such {\bc} are related to the results presented here by
elementary geometric considerations.

Now consider singular parts of free energy densities that exhibit scaling.
For the isotropic model, we assume
\begin{align}
{L'}^df_\text{iso,s}(t,L')=\Fcal_\text{iso}(\xt'),
\end{align}
with the scaling variable $\xt'\equiv(L'/\xipz)^{1/\nu}t$.
With the critical behavior (\ref{xi.crit.L}) of the correlation length
$\xi_L$ in the direction $\Lb$ of \mbox{(anti-})periodicity of the
original system, the scaling variables $\xt$ from (\ref{xt}) and $\xt'$
from above are identical, since with (\ref{Xibp.Xib}) and (\ref{Lbp})
we have
\begin{align}
\label{xtp}
(L/\xi_L)^2
&=\Lb^T\Xib^{-1}\Lb
={\Lb'}^T{\Xib'}^{-1}\Lb'
=(L'/\xi)^2.
\end{align}
Therefore, we may write for the free energy per correlation volume
\begin{align}
\label{f.per.cv.poa}
\fedt{s}(t,L)
&=
\Vco\fed{s}(t,L)=\Vco f_\text{iso,s}(t,L')
\nn\\
&=
\begin{cases}\ds
\xt^{-d\nu}\Fcal_\text{iso}(\xt)
& \mbox{for $T>\Tc$},
\\[1ex]\ds
R_\xi^{-d}(-\xt)^{-d\nu}\Fcal_\text{iso}(\xt)
& \mbox{for $T<\Tc$},
\end{cases}
\end{align}
where we have used Eq.~(\ref{Vco}) and that our shear transformation is
volume-conserving and thus separately conserves the correlation volume and
the free energy density.
Analogous equations hold for $\fed{ex,s}$, $\fed{sf,s}$, and $\fed{fs,s}$,
if they exhibit scaling.
Thus we find for $d$-dimensional films with periodic or antiperiodic {\bc}
that the leading singular parts of (scaling) free energy densities per
correlation volume depend only on the ratio of the length of
\mbox{(anti-})periodicity and the correlation length in the
corresponding direction and not on any other details of the shape or
orientation of the correlation ellipsoid.
For the two-dimensional Ising model this is reflected by the explicit
results (\ref{ffsco})--(\ref{fexco}).

Multiplying (\ref{f.per.cv.poa}) by
\begin{align}
L^d/\Vco
&=
(L/L')^d\times
\begin{cases}
\xt^{d\nu}
& \mbox{for $T>\Tc$},
\\
R_\xi^d(-\xt)^{d\nu}
& \mbox{for $T<\Tc$},
\end{cases}
\end{align}
and observing
\begin{align}
L'
&=
L|\Mb\nhb|
=
L(\nhb^T\Abar^{-1}\nhb)^{1/2},
\end{align}
which follows from (\ref{Ab}), (\ref{Lb}), and (\ref{Lbp}), we
obtain the standard free energy scaling function
\begin{align}
\label{f.Fiso.pa}
L^d\fed{s}(t,L)
&=
\Fcal(\xt,\Abar,\nhb)
\nn\\
&=
(\nhb^T\Abar^{-1}\nhb)^{-d/2}\Fcal_\text{iso}(\xt).
\end{align}
That is, $\Fcal$ depends only through a geometric factor on the relative
orientations of the direction of \mbox{(anti-)}periodicity and the
correlation ellipsoid.

Specializing to $d\,{=}\,2$--dimensional systems with conventions
as in (\ref{nhat})--(\ref{Abar.inv}), Eq.\ (\ref{f.Fiso.pa}) reads
\begin{align}
L^2\fed{s}(t,L)
&=
\Fcal(\xt,\ar,\th)
\nn\\
&=
(r\cos^2\th+r^{-1}\sin^2\th)^{-1}\Fcal_\text{iso}(\xt).
\end{align}
Analogous equations hold for $\fed{ex,s}$, $\fed{fs,s}$, and $\fed{sf,s}$,
if they exhibit scaling.
This explains, from a more general point of view, the anisotropy
dependence of the Ising model scaling functions presented in Eqs.~(\ref{Gcal.pa.3c})--(\ref{Gcal.pa}).

\subsubsection{Fixed and free \bc}
\label{d-dim.film.fixed.bc}
Here we repeat the considerations of the preceding section for a film
with fixed and/or free \bc, such as 00, ++, 0+, or +- \bc.
The treatment immediately extends to similar other invariant {\bc}.
For definiteness, we choose coordinates where the ``1'' direction is
normal to the film boundaries, i.e., where
\begin{align}
\label{nt}
\nhb^T=(1,0,\ldots,0).
\end{align}
As opposed to the case of periodic or antiperiodic \bc, the thickness of
the isotropic film is no longer given by the length of the transformed
vector $\Lb'$, see Fig~\ref{shear.trafo}.
Instead, according to Eq.~(2.48) of Ref.~\cite{KaDo10}, the film thickness
$\Lbar$ of the isotropic system is given by
\begin{align}
\label{Lbar.L}
\Lbar=(\det\Abar^{-1}/\det[[\Abar^{-1}]])^{1/2}L,
\end{align}
where $[[\Abar^{-1}]]$ is the $(d{-}1)\times(d{-}1)$ right lower part of
$\Abar^{-1}$ and where, with the  conventions employed here,
$\det\Abar^{-1}=1$, compare Eq.\ (\ref{det.A}).

For the isotropic model, we assume
\begin{align}
\Lbar^d\fed{iso,s}(t,\Lbar)=\Fcal_\text{iso}(\xbar),
\end{align}
with the scaling variable $\xbar\equiv(\Lbar/\xipz)^{1/\nu}t$.
From (\ref{Vco}), (\ref{Ab})--(\ref{Lb}), (\ref{nt}), and (\ref{Lbar.L}),
we obtain
\begin{align}
\label{L.over.xiL.sq}
(L/\xi_L)^2
&=
\Lb^T\Xib^{-1}\Lb
=
(\Abar^{-1})_{11}\det[[\Abar^{-1}]](\Lbar/\xi)^2.
\end{align}
With $\xbar$ from above and $\xt$ from (\ref{xt}) this translates to the
relation
\begin{align}
\label{xbar.xt}
\xbar
=\{(\Abar^{-1})_{11}\det[[\Abar^{-1}]]\}^{-1/(2\nu)}\xt.
\end{align}

For the singular part of the free energy per correlation volume, we may
write
\begin{align}
\label{f.per.cv.fof}
\fedt{s}(t,L)
&=
\Vco\fed{s}(t,L)=\Vco f_\text{iso,s}(t,\Lbar)
\nn\\
&=
\begin{cases}\ds
\mbox{$\xbar$}^{-d\nu}\Fcal_\text{iso}(\xbar)
& \mbox{for $T>\Tc$},
\\[1ex]\ds
R_\xi^{-d}(-\xbar)^{-d\nu}\Fcal_\text{iso}(\xbar)
& \mbox{for $T<\Tc$},
\end{cases}
\end{align}
where we have used Eq.~(\ref{Vco}) and that our shear transformation is
volume-conserving and thus separately conserves the correlation volume and
the free energy density.
Analogous equations hold for $\fed{ex,s}$, $\fed{sf,s}$, and $\fed{fs,s}$,
if they exhibit scaling.
Note that as for the case of \mbox{(anti-)}periodic \bc, the free energy
per correlation volume depends only on one suitably chosen scaling
variable.
For the two-dimensional Ising model with free \bc, this is reflected by
the explicit result (\ref{ffsco.00bc}).

Multiplying (\ref{f.per.cv.fof}) by
\begin{align}
L^d/\Vco
&=
(L/\Lbar)^d\times
\begin{cases}
\xbar^{d\nu}
& \mbox{for $T>\Tc$},
\\
R_\xi^d(-\xbar)^{d\nu}
& \mbox{for $T<\Tc$},
\end{cases}
\end{align}
and observing (\ref{det.A}) and (\ref{Lbar.L}), we obtain the standard
free energy scaling function
\begin{align}
\label{f.Fiso.ff}
L^d\fed{s}(t,L)
&=
\Fcal(\xt,\Abar,\nhb)
\nn\\
&=
(\det[[\Abar^{-1}]])^{d/2}\Fcal_\text{iso}(\xbar),
\end{align}
with $\xbar$ from (\ref{xbar.xt}).
Note that the right hand sides of Eqs.\ (\ref{Lbar.L}),
(\ref{L.over.xiL.sq}), (\ref{xbar.xt}), and (\ref{f.Fiso.ff}) change their
form for a choice of coordinates that does not imply Eq.\ (\ref{nt}).
The general form may be obtained by replacing
\bse
\begin{align}
(\Abar^{-1})_{11}&\to\nhb^T\Abar^{-1}\nhb,
\\
\det[[\Abar^{-1}]]
&\to-\f{\det(\Abar^{-1}-\nhb\nhb^T\Abar^{-1}-\Abar^{-1}\nhb\nhb^T)}%
{\nhb^T\Abar^{-1}\nhb}.
\end{align}
\ese

Specializing to $d\,{=}\,2$--dimensional systems with conventions
as in (\ref{nhat})--(\ref{Abar.inv}), Eq.\ (\ref{f.Fiso.ff}) reads
\begin{align}
\label{f.Fiso.2d.ff}
L^2\fed{s}(t,L)
&=
\Fcal(\xt,\ar,\th)
\nn\\
&=
(r\sin^2\th+r^{-1}\cos^2\th)\Fcal_\text{iso}(\xbar),
\end{align}
with
\begin{align}
\label{xbar.xt.2}
\xbar
&=
\left[1+\fr{1}{4}(r-r^{-1})^2\sin^2(2\th)\right]^{-1/(2\nu)}\xt,
\end{align}
as obtained by combining $(\Abar^{-1})_{11}$ and
$\det([[\Abar^{-1}]])=(\Abar^{-1})_{22}$ from (\ref{Abar.inv}) with
(\ref{xbar.xt}).
With $\nu$ from (\ref{nu.xi0p.xi0m}), Eq.\ (\ref{xbar.xt.2}) reduces to
the relation (\ref{xbar.Ising}) found for the two-dimensional Ising model.
Equations analogous to (\ref{f.Fiso.2d.ff}) hold for $\fed{ex,s}$,
$\fed{sf,s}$, and $\fed{fs,s}$, if they exhibit scaling.
As for periodic and antiperiodic \bc, the $d$--dimensional point of view
provides an explanation for the anisotropy dependence of the Ising model
scaling function for free {\bc} presented in (\ref{G00}).

In Appendix~\ref{iso.scal.func.ff}, we sketch the derivation of the
isotropic Ising model results
\bse
\label{Gcal.fixed}
\begin{align}
\label{Gcal.++.00}
\Gcaliso^\ppbc(\xt)
&=
\Gcaliso^\zzbc(-\xt)
=\If(-\xt),
\\
\label{Gcal.pz.Gcal.a}
\Gcaliso^\zpbc(\xt)
&=
\Gcaliso^\abc(2\xt)/4
=\Im(2\xt)/4,
\end{align}
\ese
with $\If$ and $\Im$ from (\ref{If}) and (\ref{Ipm}), respectively.
With $x=2\xt$, our result for $\Gcaliso^\ppbc(\xt)$ is identical to
$X_\text{ex}^{(+,+)}(x)$ provided at the beginning of Sec.~12.1.2 in
\cite{BrDaTo00} without derivation.
$\Gcaliso^\ppbc(\xt)$ and $\Gcaliso^\zpbc(\xt)$ are shown in
Fig.~\ref{Gisofig}.
Together with (\ref{xbar.Ising}), (\ref{f.per.cv.fof}), and
(\ref{f.Fiso.2d.ff}), we immediately obtain predictions for both the
finite-size scaling behavior of the finite-size part of the free energy
per correlation volume and the scaling functions of the finite-size part
of the free energy for the anisotropic case for both $++$ and $0+$ {\bc}.
Combining (\ref{f.Fiso.2d.ff}) with (\ref{Gcal.fixed}), while
observing (\ref{Ipm.0}) and (\ref{If.0}), the related Casimir amplitudes
are
\begin{align}
\label{De.++.00.0+}
\De^\ppbc(r,\th)=-\De^\zpbc(r,\th)/2&=\De^\zzbc(r,\th),
\end{align}
with $\De^\zzbc(r,\th)$ from (\ref{De00}).
With the definition (\ref{De.INW}), the results provided at the end of
Sec.~2 in Ref.~\cite{InNiWa86} for the Casimir amplitudes are, for ++, 0+,
and 00 {\bc},
\begin{align}
\label{De.INW.++.00.0+}
\De_\text{INW}^\ppbc
&=-\De_\text{INW}^\zpbc/2
=\De_\text{INW}^\zzbc,
\end{align}
with $\De_\text{INW}^\zzbc$ from (\ref{De.INW.00}).
While the results for $\De_\text{INW}^\ppbc$ and $\De_\text{INW}^\zpbc$
were derived in Ref.~\cite{InNiWa86} only for rectangular lattices, the
combination of (\ref{DeINW.De}), (\ref{De.++.00.0+}), and
(\ref{De.INW.++.00.0+}) shows that they remain correct for the general
triangular lattice.

\section{Summary and discussion}
\label{summary}
We have investigated finite-size scaling of an anisotropic two-dimensional
ferromagnetic Ising model on an infinite strip of width $L$ with periodic,
antiperiodic, and free {\bc} in the direction perpendicular to the
direction of infinite extent.
The model is realized on a triangular lattice with general couplings and
lattice constants, so that the anisotropy may be realized by varying the
lattice constants $d_i$ and/or the couplings $J_i$.
This allows for different microscopic realizations of identical
anisotropic bulk long-distance correlations near the bulk critical point
and therefore for a limited test of universality.

We find that the asymptotic scaling behavior of the finite-size
contribution to the free energy per correlation volume may be described by
a function of only one suitably defined scaling variable, cf.\ Eqs.\
(\ref{ffsco}) and (\ref{ffsco.00bc}).
For periodic and antiperiodic {\bc} this holds additionally for the
surface and excess free energy densities per correlation volume, cf.\
Eqs.\ (\ref{fsfco}) and (\ref{fexco}), while, for free {\bc}, logarithmic
violations of scaling prevent the scaling behavior of the surface
contribution to the free energy density.

For periodic, antiperiodic, and free {\bc}, we provide exact scaling
functions $\Gcal^\pbc$, $\Gcal^\abc$, and $\Gcal^\zzbc$ of the
finite-size contribution to the free energy density in Eqs.\
(\ref{Gcal.pa.3c}) and (\ref{Gcal.00}) and scaling functions $\Acal^\pbc$
and $\Acal^\abc$ of the excess free energy density in
Eq.~(\ref{Acal.pa.3c}).
We find that these functions only depend on variables related to
long-distance correlations.
We have chosen these long-distance variables to be the scaling variable
$\xt$ representing the ratio of the width $L$ of the strip and the bulk
correlation length in the corresponding direction (for $T>\Tc$; for
$T<\Tc$ its analytic continuation is employed) as well as the aspect ratio
$\ar$ and the orientation angle $\th$ of the bulk correlation lengths
ellipse.
Consequently, the critical Casimir amplitudes $\De^\pbc$, $\De^\abc$, and
$\De^\zzbc$ provided in Eqs.\ (\ref{De.p.a}) and (\ref{De00}) depend only
on $\ar$ and $\th$.
Since our results for $\Gcal$, $\Acal$, and $\De$ are independent of the
microscopic realization of the long-distance physics, we conjecture that
they are universal functions of their respective arguments.
That is, we expect identical functions to result for other members of the
bulk two-dimensional Ising universality class for the same geometry and
corresponding \bc.

To understand the behaviors of the Ising systems described above, we have
investigated general anisotropic systems on a $d$-dimensional film with
the same {\bc} as investigated for the two-dimensional Ising model.
We assume that the free energies under consideration exhibit scaling.
Due to the choice of geometry and \bc, these systems have the simplifying
property that their geometry and {\bc} are invariant under a shear
transformation relating them to a corresponding isotropic system.
We find that the free energies per correlation volume of such systems
depend on only one suitably chosen scaling variable, cf.\
Eqs.\ (\ref{f.per.cv.poa}) and (\ref{f.per.cv.fof}).
The relations between the scaling functions of the isotropic and the
anisotropic systems are provided in Eqs.\ (\ref{f.Fiso.pa}) and
(\ref{f.Fiso.ff}) and reproduce for $d=2$ the relations found for the
two-dimensional Ising model.
For $++$ and $0+$ \bc, where no results for the triangular lattice Ising
model are available, we point out that these relations, together with the
explicitly known isotropic results (\ref{Gcal.fixed}), lead to predictions
for the corresponding anisotropic scaling functions of the finite-size
contribution to the free energy density of the two-dimensional Ising
model.
We give explicit results for the corresponding critical Casimir
amplitudes in (\ref{De.++.00.0+}).

We have only treated geometries and {\bc} that are invariant under the
shear transformation that relates the system under investigation to a
system with isotropic bulk correlation lengths. 
In other cases we cannot expect to express scaling functions for the
anisotropic case in terms of an isotropic-case scaling function with the
same geometry and \bc.
Rather, the isotropic scaling functions will depend on the geometry and
{\bc} obtained by transforming the anisotropic to the isotropic case
and the anisotropy will no longer be represented by mere geometric
factors as in (\ref{Gcal.pa.3c})--(\ref{Acal.pa.3c}), (\ref{xbar.Ising}),
(\ref{Gcal.00}), (\ref{f.Fiso.pa}), (\ref{xbar.xt}), and (\ref{f.Fiso.ff}).
However, we still expect such scaling functions to be universal functions
of their arguments.
Within a given bulk universality class, they should depend only on the
scaling variable(s) used for the isotropic case, the asymptotic
long-distance anisotropy represented by the matrix $\Abar$ from
(\ref{Ab}), the geometry, and the \bc.
Hence we suggest that a quantity is universal, if it depends only on the
{\bc} and on macroscopic physical observables such as the geometry and the
macroscopic near-critical correlation lengths, but not on the particular
microscopic realization from which the anisotropic long-distance critical
behavior originates.
Such a quantity should therefore be identical among the members of the
bulk universality class under consideration.
For practical measurements of, e.g., scaling functions, it is no longer
sufficient to measure one correlation length as in the isotropic case.
Rather, the measurement of correlation lengths in a sufficient number of
directions is necessary to allow for the determination of the matrix
$\Abar$ through (\ref{Ab}).

Let us put our arguments in perspective with the interpretations of
\cite{ChDo04,Do06,Do08,DiCh09}.
From our arguments follows that it is not necessary to define universality
only after a shear transformation to an isotropic system, as Diehl and
Chamati suggest \cite{DiCh09}.
As noted in Sec.~\ref{corr.len.film}, Chen and Dohm
\cite{ChDo04,Do06,Do08} define their matrix $\Abar$ through microscopic
parameters in the Hamiltonian so that, at least for lattice models, their
matrix is only an approximation to our definition which relates $\Abar$
through (\ref{Ab}) to the asymptotic physical correlation lengths.
Thus the dependence of scaling functions on their matrix $\Abar$ will, in
general, not be universal, since it depends on the microscopic realization
of the anisotropy.
In this sense it is correct when Chen and Dohm note that any dependence
of physical quantities on their anisotropy matrix $\Abar$ is nonuniversal
\cite{ChDo04,Do06,Do08}.
In contrast, we suggest here that quantities that are universal for the
isotropic case, merely acquire an additional universal dependence on
$\Abar$ from (\ref{Ab}) for the anisotropic case.
The parameters describing $\Abar$ are called nonuniversal in \cite{Do08}.
This is correct in the sense that the relation of $\Abar$ from
\cite{ChDo04,Do06,Do08} to our $\Abar$ depends on the model under
consideration. 
However, with our definition, $\Abar$ is merely an argument of, e.g., a
scaling function, and should be viewed on the same level as the scaling
variable $\xt$.
It is then only the scaling functions, whose universality can be tested
and not that of their arguments $\xt$ and $\Abar$.
Similar arguments hold for the parameters describing the shear
transformation, which are called nonuniversal in \cite{Do08,DiCh09}.
This transformation merely describes a relation between different physical
situations (or different interpretations of the same statistical model),
not necessarily one of them being isotropic.
With our definitions, its classification as universal or nonuniversal
is not a meaningful question.

Our last argument concerns the validity of two-scale factor universality.
Since $\Abar$ is a scale-free quantity, our universality interpretation
does not interfere with two-scale factor universality in a formulation
including the field $h$ conjugate to the order parameter.
For definiteness, consider again film geometry with given {\bc} and assume
that the free energy density exhibits scaling.
Then we expect the universal isotropic scaling function
\cite{PrFi84,Pr90,PrAhHo91}
\begin{align}
L^df_s(t,h,L)=\Fcal(C_1tL^{1/\nu},C_2hL^{\be\de/\nu}),
\end{align}
with nonuniversal scale factors $C_1$ and $C_2$ and critical exponents
$\be$ and $\de$ to be generalized to an anisotropic universal scaling
function
\begin{align}
\label{Fcal.h.aniso}
L^df_s(t,h,L)
&=
\Fcal(C_1tL^{1/\nu},C_2hL^{\be\de/\nu},\Abar,\nhb),
\end{align}
where the choice $C_1={\xipz^{(L)}}^{-1/\nu}$ makes the first argument
identical to $\xt$ in (\ref{xt}) and leads to the $h{=}0$ limit
$\Fcal(\xt,0,\Abar,\nhb)=\Fcal(\xt,\Abar,\nhb)$, compare
(\ref{Fcal.aniso}).
Since no new nonuniversal scale factor needed to be introduced into
Eq.~(\ref{Fcal.h.aniso}) as compared to the isotropic case, two-scale
factor universality remains valid.
We have not quantitatively considered this case, since no exact results
for the free energy of the two-dimensional Ising model with a nonzero
magnetic field are available.

As an outlook, it would be interesting to investigate systems that do not
exhibit the simplifying feature of both invariant geometry and {\bc} under
shear transformations.
A possible example is the critical Binder cumulant for a two-dimensional
ferromagnetic Ising model.
Precision results for this quantity are available for a square lattice in
a square geometry with periodic {\bc} and a $45^\circ$ anisotropy that is
caused by a ferromagnetic \cite{KaBl93,SeSh05} or antiferromagnetic
\cite{SeSh09} coupling on one of the lattice diagonals, i.e., by
different microscopic realizations.
For recent results of such an investigation, see Ref.~\cite{Ka12}.

\section*{Acknowledgments}
The author is grateful to V.~Dohm for suggesting this investigation and
for helpful discussions and correspondence.

\appendix

\section{Integrals and sums}
Here we collect some mathematical results needed for the determination of
the scaling functions of the free energy in Sec.~\ref{free.energy.pa}.

On the one hand, we need, for sufficiently well-behaved functions $f$, the
large-$n$ results
\bse
\label{Iabml}
\begin{align}
\label{Iabm}
\int_a^bdxf(x)
=&
\sum_{j=1/2}^{n-1/2}f(x_j)\de
+\f{1}{24}[f'(b)-f'(a)]\de^2+\Ocal(\de^3),
\\
\label{Iabl}
\int_a^bdxf(x)
=&
\sum_{j=0}^{n-1}f(x_j)\de
+\f{1}{2}[f(b)-f(a)]\de
\nn\\
&
-\f{1}{12}[f'(b)-f'(a)]\de^2+\Ocal(\de^3),
\end{align}
\ese
respectively, where $\de\equiv(b-a)/n$ and $x_j\equiv a+j\de$.

On the other hand, using the method of residues, we obtain in the
large-$n$ limit for real $x$
\bse
\label{delta.sq.pa}
\begin{align}
\label{delta.sq.p}
&\f{1}{\pi}\int_0^{n\pi}d\om\,
\sqrt{x^2+4\om^2}
-\sum_{k=0}^{n-1}\sqrt{x^2+[(2k+1)\pi]^2}
\nn\\
&=
\f{\pi}{12}+\Ip(x),
\\
\label{delta.sq.a}
&\f{1}{\pi}\int_0^{(n-\f{1}{2})\pi}d\om\,\sqrt{x^2+4\om^2}
-\sum_{k=1}^{n-1}\sqrt{x^2+(2k\pi)^2}
\nn\\
&=
\f{|x|}{2}-\f{\pi}{6}+\Im(x),
\end{align}
\ese
with $\Ipm$ defined in (\ref{Ipm}).

The results (\ref{Iabml}) and (\ref{delta.sq.pa}) allow us to obtain Eq.~(\ref{fex.pa.raw.tr}) from Eq.~(\ref{fex.pa.int.sum.tr}).

\section{Isotropic scaling functions}
\label{iso.scal.func}
Here we relate the scaling functions $\Gcal$ and $\Acal$ to published
scaling functions of the Casimir force for the isotropic two-dimensional
Ising model in infinite-strip geometry.
This allows us to verify the isotropic limits of our results for
$\Gcal^\pbc$, $\Gcal^\abc$, and $\Gcal^\zzbc$ and to derive the
scaling functions $\Gcaliso^\zzbc$ and $\Gcaliso^\zpbc$ that can
subsequently be used for predictions about the anisotropic case as
explained in Sec.~\ref{d-dim.film.fixed.bc}.

For the isotropic case, the Casimir force of a $d$-dimensional film system
of thickness $L$ is defined by
\begin{align}
\label{FCas.def}
F_\text{Cas,iso}(T,L)
&\equiv
-\f{\p[L\fed{iso,ex}(T,L)]}{\p L}
\nn\\
&=
-\f{\p[L\fed{iso,fs}(T,L)]}{\p L},
\end{align}
where the second equality reflects the fact that surface contributions do
not contribute to the Casimir force.
If its singular part $F_\text{Cas,iso,s}$ exhibits scaling, a
corresponding scaling function $\Xiso$ may be defined in the asymptotic
critical domain, so that
\begin{align}
L^dF_\text{Cas,iso,s}&(t,L)
=
\Xiso(\xt)
\nn\\
&=
(d-1)\Gcaliso(\xt)-\nu^{-1}\xt\f{d\Gcaliso(\xt)}{d\xt},
\end{align}
where (\ref{Gcal}) and (\ref{xt}) have been used.
With (\ref{nu.xi0p.xi0m}), this may be expressed, for the two-dimensional
Ising model, as
\begin{align}
\label{FcasCas.def}
L^2F_\text{Cas,iso,s}(t,L)=\Xiso(\xt)
=-\xt^2\f{d[\Gcaliso(\xt)/\xt]}{d\xt}.
\end{align}

\subsection{Periodic and antiperiodic \bc}
\label{iso.scal.func.pa}
Here we use results from the literature for $\Xiso^\psabc$ to verify the
isotropic limits (\ref{Gcal.iso.pa}) of the scaling functions
$\Gcal^\psabc$ provided in (\ref{Gcal.pa.3c}).
Rescaling $\om\to|x|\om$ in (\ref{Ipm}), applying (\ref{FcasCas.def})
to the resulting expression for $\Gcaliso^\psabc(\xt)$ and
subsequently scaling back according to $\om\to\om/|x|$ gives
\begin{align}
\label{Xpsa}
\Xiso^\psabc(\xt)
&=
\f{1}{2\pi}
\int_0^\infty d\om
\sqrt{\xt^2+\om^2}
\nn\\
&\ph{=}
\times
\begin{cases}
\left[\tanh\left(\sqrt{\xt^2+\om^2}/2\right)-1\right]
\\[2ex]
\left[\coth\left(\sqrt{\xt^2+\om^2}/2\right)-1\right]
\end{cases},
\end{align}
where the upper and lower expressions hold for periodic and antiperiodic
{\bc}, respectively.
Corresponding critical values are
\begin{align}
\label{Xpsa.0}
\Xiso^\psabc(0)
&=
\begin{cases}
-\pi/12,
\\[0.5ex]
\pi/6.
\end{cases}
\end{align}

The upper result in (\ref{Xpsa}) agrees numerically
with the Ising curve plotted in Fig.~15 of Ref.~\cite{VaGaMaDi09}
($x$ there is identical to our $\xt$).
The upper and lower results in (\ref{Xpsa}) agree with Eqs.~(3.16) and
(3.19), respectively, of Ref.~\cite{RuZaShAb10}
($x$ there is identical with our $\xt/2$).
The upper result in (\ref{Xpsa.0}) agrees with Eq.~(1) in
Ref.~\cite{BlCaNi86}, with Eq.~(3) in Ref.~\cite{Af86}, and with Eq.~(58)
in Ref.~\cite{HuGrSc10}.

\subsection{Fixed and free \bc}
\label{iso.scal.func.ff}
Here we use results from Refs.~\cite{EvSt94} and \cite{AbMa10} for the
Casimir force scaling functions $\Xiso^\zzbc$, $\Xiso^\ppbc$, and
$\Xiso^\zpbc$, for the isotropic two-dimensional Ising model in
infinite-strip geometry to derive the scaling functions $\Gcaliso^\zzbc$,
$\Gcaliso^\ppbc$, and $\Gcaliso^\zpbc$.
From Eqs.~(2.9) and (2.13) in Ref.~\cite{EvSt94} follows for the
isotropic-case Casimir force scaling functions for ++ and 00 {\bc} in our
notation ($X$ there is identical to our $\xt/2$)
\begin{align}
\label{FcalCas.++.00}
\Xiso^\ppbc(\xt)
&=
\Xiso^\zzbc(-\xt)
\nn\\
&=
-\f{1}{\pi}\int_0^\infty \f{d\om\sqrt{\xt^2+\om^2}}%
{1+\f{\sqrt{\xt^2+\om^2}-\xt}{\sqrt{\xt^2+\om^2}+\xt}
e^{2\sqrt{\xt^2+\om^2}}}.
\end{align}
Equivalent results are obtained from Eq.~(3) in Ref.~\cite{AbMa10} for
both 00 and ++ {\bc} ($x$ there is identical to our $\xt$)
\cite{comment_AbMa10} and from Eq.~(3.26) in Ref.~\cite{RuZaShAb10} for 00
{\bc} ($x$ there is identical to our $\xt/2$), while the results from
Sec.~12.1.2 of \cite{BrDaTo00} for $X_\text{Cas}^{(o,o)}(x)$ and
$X_\text{Cas}^{(+,+)}(x)$ ($x$ there is our $2\xt$) are missing a factor
$1/4$.
Using (\ref{FcasCas.def}) and observing that, for large positive or
negative $\xt$, the scaling function $\Gcal$ contains by definition no
surface or interface terms, i.e., no terms linear in $\xt$, we obtain from
(\ref{FcalCas.++.00}) by elementary integration the scaling functions
$\Gcaliso^\zzbc$ provided in (\ref{G00.iso}) and (\ref{Gcal.++.00}) and
$\Gcaliso^\ppbc$ provided in (\ref{Gcal.++.00}) .

From Eq.~(3) in Ref.~\cite{AbMa10} ($x$ there is identical to our $\xt$),
we obtain \cite{comment_AbMa10}
\begin{align}
\label{FcalCas.pz}
\Xiso^\zpbc(\xt)
&=
\f{1}{2\pi}\int_0^\infty d\om\sqrt{\xt^2+\om^2}
\left(\coth\sqrt{\xt^2+\om^2}-1\right)
\nn\\
&=
\Xiso^\abc(2\xt)/4,
\end{align}
with $\Xiso^\abc$ from (\ref{Xpsa}).
With (\ref{FcasCas.def}) and taking again into account the absence of
large-$|\xt|$ linear terms in $\Gcal$, elementary integration leads
immediately to $\Gcal^\zpbc$ as provided in (\ref{Gcal.pz.Gcal.a}).


\begin{thebibliography}{99}
\bibitem{comment_InNiWa86_a}
The minus sign is missing in Eq.~(A22) of Ref.~\cite{InNiWa86}.
\bibitem{comment_LiHuLiWu99}
The minus sign in Eq.~(66) of Ref.~\cite{LiHuLiWu99} needs to be replaced
by a plus sign.
\bibitem{comment_InNiWa86_b}
The ``$\ln$'' is missing in the second term on the right hand side of
Eq.~(A15) of Ref.~\cite{InNiWa86}.
No expression for $\bar{a}$ was provided in Ref.\ \cite{InNiWa86}.
We have simplified the expressions for $p_\pm$.
\end{thebibliography}
\end{document}